\documentclass[fleqn,usenatbib]{mnras}

\usepackage{newtxtext,newtxmath}
\usepackage[caption=false]{subfig}

\usepackage[T1]{fontenc}

\DeclareRobustCommand{\VAN}[3]{#2}
\let\VANthebibliography\thebibliography
\def\thebibliography{\DeclareRobustCommand{\VAN}[3]{##3}\VANthebibliography}

\usepackage{graphicx}	
\usepackage{amsmath}	
\usepackage{verbatim}
\usepackage{multirow}
\usepackage{multicol}

\usepackage{comment}
\usepackage{xspace}
\usepackage{enumitem}
\usepackage{graphicx}
\usepackage{rotating}
\usepackage{color}

\newcommand{\oii}{[O\,{\sc ii}]}

\newcommand{\chisq}{$\chi^2_{\nu}$}

\newcommand{\fc}{$\rm f_{\rm c}$}

\newcommand{\ergscmarc}{$\rm erg\,s^{-1}\,cm^{-2}\,arcsec^{-2}$}
\newcommand{\mstar}        {\relax\ifmmode{{M}_{\rm *}\xspace} \else {${M}_{\rm *}$}\expandafter\xspace\fi}
\newcommand{\msun}        {\relax\ifmmode{{M}_{\odot}\xspace} \else {${M}_{\odot}$}\expandafter\xspace\fi}

\newcommand{\lya}        {Ly$\alpha$\xspace}
\newcommand{\hi}         {\ion{H}{I}\xspace}
\newcommand{\mgii}         {\ion{Mg}{II}\xspace}

\newcommand{\rp}        {\relax\ifmmode{{ R}_{\rm p}\xspace} \else {${ R}_{\rm p}$}\expandafter\xspace\fi}

\newcommand{\fmghi}        {\relax\ifmmode{{ f}_{\rm MgII/HI}\xspace} \else {${ f}_{\rm MgII/HI}$}\expandafter\xspace\fi}
\newcommand{\NHI}        {\relax\ifmmode{{ N}_{\rm HI}\xspace} \else {${ N}_{\rm HI}$}\expandafter\xspace\fi}
\newcommand{\Nmg}        {\relax\ifmmode{{ N}_{\rm MgII}\xspace} \else {${ N}_{\rm MgII}$}\expandafter\xspace\fi}
\newcommand{\tauK}        {\relax\ifmmode{{ \tau}_{\rm K}\xspace} \else {${ \tau}_{\rm K}$}\expandafter\xspace\fi}
\newcommand{\tauH}        {\relax\ifmmode{{ \tau}_{\rm H}\xspace} \else {${ \tau}_{\rm H}$}\expandafter\xspace\fi}

\newcommand{\lambdaK}        {\relax\ifmmode{{ \lambda}_{\rm K}\xspace} \else {${ \lambda}_{\rm K}$}\expandafter\xspace\fi}
\newcommand{\lambdaH}        {\relax\ifmmode{{ \lambda}_{\rm H}\xspace} \else {${ \lambda}_{\rm H}$}\expandafter\xspace\fi}

\newcommand{\kms}        {\ifmmode{\rm \,km\,s^{-1}}\else\,km\,s$^{-1}$\xspace\fi}
\newcommand{\unitNHI}    {\ifmmode{\rm \,cm^{-2}}\else\,cm$^{-2}$\xspace\fi}  
\newcommand{\vexp}       {\relax\ifmmode {v_{\rm exp}} \else {$v_{\rm exp}$}\expandafter\xspace\fi}
\newcommand{\vran}       {\relax\ifmmode {v_{\rm ran}} \else {$v_{\rm ran}$}\expandafter\xspace\fi}
\newcommand{\vth}       {\relax\ifmmode {v_{\rm th}} \else {$v_{\rm th}$}\expandafter\xspace\fi}
\newcommand{\sigran}     {\relax\ifmmode {\sigma_{\rm Ran}} \else {$\sigma_{\rm Ran}$}\expandafter\xspace\fi}
\newcommand{\sigr}     {\relax\ifmmode {\sigma_{\rm R}} \else {$\sigma_{\rm R}$}\expandafter\xspace\fi}
\newcommand{\sigsrc}     {\relax\ifmmode {\sigma_{\rm Src}} \else {$\sigma_{\rm Src}$}\expandafter\xspace\fi}
\newcommand{\sigconv}     {\relax\ifmmode {\sigma_{\rm Conv}} \else {$\sigma_{\rm Conv}$}\expandafter\xspace\fi}

\newcommand{\fsim}       {\relax\ifmmode {f_{\rm sim}} \else {$f_{\rm sim}$}\expandafter\xspace\fi}
\newcommand{\fobs}       {\relax\ifmmode {f_{\rm obs}} \else {$f_{\rm obs}$}\expandafter\xspace\fi}
\newcommand{\zobs}       {\relax\ifmmode {z_{\rm obs}} \else {$z_{\rm obs}$}\expandafter\xspace\fi}

\newcommand{\EWemis}        {\relax\ifmmode{|{ \rm EW}_{\rm emis}|\xspace} \else {$|{ \rm EW}_{\rm emis}|$}\expandafter\xspace\fi}
\newcommand{\EWabs}        {\relax\ifmmode{|{ \rm EW}_{\rm abs}|\xspace} \else {$|{ \rm EW}_{\rm abs}|$}\expandafter\xspace\fi}
\newcommand{\EWint}        {\relax\ifmmode{{ \rm EW}_{\rm int}\xspace} \else {${ \rm EW}_{\rm int}$}\expandafter\xspace\fi}
\newcommand{\EWobs}        {\relax\ifmmode{|{ \rm EW}_{\rm obs}|\xspace} \else {$|{ \rm EW}_{\rm obs}|$}\expandafter\xspace\fi}

\title[Modeling \mgii Emission from SFG at z $\sim$ 1]
{Modeling \mgii resonance doublet spectra from galaxy haloes at z $\sim$ 1}

\author[S.-J. Chang et al.]{
Seok-Jun Chang,$^{1}$\thanks{E-mail: sjchang@mpa-garching.mpg.de}
Rajeshwari Dutta,$^{2}$
Max Gronke,$^{1}$
Michele Fumagalli,$^{3,4}$ 
Fabrizio Arrigoni Battaia,$^{1}$
\newauthor
Matteo Fossati,$^{3,5}$
\\
$^{1}$ Max-Planck-Institut f\"{u}r Astrophysik, Karl-Schwarzschild-Stra$\beta$e 1, 85748 Garching b. M\"{u}nchen, Germany \\
$^{2}$ IUCAA, Postbag 4, Ganeshkind, Pune 411007, India \\
$^{3}$ Dipartimento di Fisica ``G. Occhialini'', 
Universit\`a degli Studi di Milano-Bicocca, 
Piazza della Scienza 3, I-20126 Milano, Italy \\
$^{4}$ INAF – Osservatorio Astronomico di Trieste, Via G. B. Tiepolo 11, I-34143 Trieste, Italy \\
$^{5}$ INAF – Osservatorio Astronomico di Brera, 
Via Brera 28, I-21021 Milano, Italy}

\date{Accepted XXX. Received YYY; in original form ZZZ}

\pubyear{2024}

\begin{document}
\label{firstpage}
\pagerange{\pageref{firstpage}--\pageref{lastpage}}
\maketitle

\begin{abstract}
We investigate the properties of cold gas at $10^4~\rm K$ around star-forming galaxies at $z~\sim~1$ using Mg~II spectra through radiative transfer modeling. We utilize a comprehensive dataset of 624 galaxies from the MAGG and MUDF programs. We focus on Mg~II emission from galaxies and their outskirts to explore the cold gas within galaxies and the circumgalactic medium (CGM). We model Mg~II spectra for 167 individual galaxies and stacked data for different stellar mass bins. The Mg~II spectrum and surface brightness vary significantly with stellar mass. In low-mass galaxies ($M_*/M_\odot<10^9$), Mg~II emission is observed in both core ($R_{\rm p}<$~10~kpc) and halo regions (10~kpc~$<R_{\rm p}<$~30~kpc), while in higher mass galaxies ($M_*/M_\odot>10^{10}$), strong core absorption and more extended halo emission are prominent. This indicates that more massive galaxies have more cold gas. Radiative transfer modeling allows us to investigate key parameters such as the Mg~II column density $N_{\rm MgII}$ and the outflow velocity $v_{\rm exp}$. We identify a negative correlation between $N_{\rm MgII}$ and $v_{\rm exp}$. Since higher stellar mass galaxies exhibit a higher $N_{\rm MgII}$ and lower $v_{\rm exp}$, this suggests an abundance of slowly moving cold gas in massive galaxies. In addition, the fitting results of halo spectra indicate the presence of intrinsic Mg~II absorption and strong anisotropy of the cold gas distribution around massive galaxies. This study is not only a proof-of-concept of modeling spatially varying Mg~II spectra but also enhances our understanding of the CGM and provides insights into the mass-dependent properties of cold gas in and around galaxies.
\end{abstract}




\section{Introduction}
\label{sec_introduction}

The interplay between galaxies and their surrounding circumgalactic medium (CGM) is a fundamental aspect of galactic evolution, influencing processes ranging from star formation to galactic inflows and outflows. 
The CGM represents a complex, multiphase structure around galaxies, acting as both a reservoir for star formation in the interstellar medium (ISM) and a repository for infalling material from the intergalactic medium (IGM) and for material ejected from galaxies by feedback processes. This diffuse halo, extending tens to hundreds of kiloparsecs and bounded by the dark matter halo, is a medium where the ISM and the IGM meet, facilitating the exchange of gas, dust, and energy. The multiphase nature of the CGM is characterized by a coexistence of gas in various states composed of a cold component, consisting of small dense clumps, and a hot component, consisting of ionized regions heated by supernovae, and active galactic nuclei \citep[see reviews][]{tumlinson17,faucher-giguere23}.

Historically, due to its diffuse nature, the CGM has been investigated primarily through absorption lines detected against background sources \citep[e.g.][]{prochaska13,crighton14,bouche16,chen20,Dutta2020,weng23}. 
While this method is effective for probing specific lines of sight, it falls short of providing a comprehensive view of the overall CGM structure. To address this limitation, recent advancements in telescopes and instrumentation have enabled studies of the CGM via emission lines, allowing us to better understand the physical properties of the gas in galaxy outskirts \citep[e.g.][]{steidel10,steidel11,hennawi13,fabrizio15,wisotzki16,fab19,leclercq22,gronzalez23,dutta23,Dutta2024,guo23,guo24,tornotti2024}.

The \mgii resonance doublet at 2796 and 2803 \AA\ is emerging as a new tracer of the cold CGM at $T\sim 10^4\, \rm K$.
While \lya emission is commonly used for studying the cold gas due to its resonance nature \citep[see reviews][]{dijkstra19,ouchi20}, and the optically thick nature of the neutral IGM, it has limitations at the epoch of reionization ($z > 7$). Moreover, at lower redshifts ($z < 2$), \lya observations require space-based telescopes.
Because of the similar ionization energy of \mgii ($\sim 15\,$eV) compared to neutral hydrogen ($13.6\,$eV), the \mgii resonance line has been established as a complementary tracer of cold gas. In addition to not being hindered by neutral hydrogen residing in the IGM, \mgii is observable from the ground at $0.1 < z < 2$.
Recent observational studies have already shown the presence of extended \mgii emission halos around galaxies at $z < 2$ \citep[][]{rubin11,rickards19,zabl21,burchett21,leclercq22,dutta23,guo23,pessa24}, and in particular highlighted \mgii as an indicator of LyC escape \citep{henry18,chisholm20,seive22,izotov22,xu23}.
Similarly, theoretical work has highlighted the potential of \mgii observables in constraining physical properties of the gas in galaxies, particularly in conjunction with other resonant lines such as \lya \citep{prochaska11,seon24,chang24}.


To fully decode the observational features of the \mgii line, understanding its radiative transfer processes is crucial. The interaction between \mgii photons and atoms in the cold gas causes both spatial and spectral variations, which are key to interpreting the observed \mgii emission \citep{prochaska11,katz22,seon24,chang24}. 
In some recent studies, the spatially extended \mgii halo are modeled using primarily absorption features imprinted on the observed spectrum \citep{zabl21,pessa24}.
Furthermore, \citet{li24} interpreted \mgii and \lya spectra of 33 LyC leakers at $z \sim 0.2$ through full radiative transfer modeling.
These previous studies show that it is possible to reproduce and interpret the observed \mgii emission, and overall highlight the importance and potential of radiative transfer modeling to investigate cold gas properties through \mgii emission.

In this paper, we continue these efforts and investigate \mgii spectra of star-forming galaxies at $z \sim 1$ through extensive radiative transfer modeling \citep[also see][]{carr2024}.
We not only model individual spectra but explore \mgii emission extending beyond 10 kpc from the galactic center via stacking. 
In \S~\ref{sec:observations}, we describe the observational data, emphasizing the characteristics that make them particularly suited for our study of the \mgii emission. 
\S~\ref{sec:modeling} introduces our radiative transfer simulations and explains the key physical parameters that influence \mgii profiles.
In \S~\ref{sec:stacking_observation}, we present the results from stacked observations across different stellar mass bins. 
In \S~\ref{sec:result}, we interpret cold gas properties based on our modeling.
\S~\ref{sec:conclusion} summarizes our findings and discusses prospects for future investigations into the CGM using \mgii.

\section{Observations}
\label{sec:observations}

\subsection{Galaxy Sample}
\label{sec:observations_sample}

We use observations from two Multi Unit Spectroscopic Explorer \citep[MUSE;][]{Bacon2010} large programs on the Very Large Telescope (VLT), namely the MUSE Analysis of Gas around Galaxies (MAGG) and the MUSE Ultra Deep Field (MUDF). The two surveys, MAGG, comprising medium-deep ($\approx5-10$\,h) MUSE observations of 28 $1\times1$ arcmin$^2$ fields centered on $z\approx3.2-4.5$ quasars, and MUDF, comprising ultra-deep ($\approx$143\,h) MUSE observations of a $1.5\times1.2$ arcmin$^2$ field with two $z\approx3.2$ quasars, are complementary in depth and volume. The spectral resolution of the MUSE data range between 1770 at 4650\,\AA\ to 3590 at 9300\,\AA, and the average image quality of the MUSE data is $\approx 0.7$ arcsec full-width at half-maximum.  The MAGG survey is described in detail in \citet{Lofthouse2020,Lofthouse2023}, \citet{Dutta2020,Dutta2021,dutta23}, \citet{Fossati2021}, and \citet{Galbiati2023}, while the details of the MUDF survey are presented in \citet{Lusso2019}, \cite{Fossati2019}, \citet{Revalski2023}, and \citet{dutta23}. The galaxy catalogs of the MAGG and MUDF surveys are described in detail in Section 2 of \citet{dutta23}. For the \mgii\ stacking discussed in this work, we follow the same procedure as described in section 3 of \citet{dutta23}, with an updated galaxy sample as explained below.

In \citet{dutta23}, we considered galaxies in MAGG and MUDF within the redshift range $0.7 \le z \le 1.5$ where there is coverage of both the \oii\ and the \mgii\ line in the MUSE spectra. This work is focused on modeling the \mgii\ line. Therefore, we consider the entire redshift range $0.7 \le z \le 2.3$ with coverage of the \mgii\ line in MUSE. The analysis in \citet{dutta23} was based on the MUSE observations of the MUDF survey that had been acquired till 2021 November 21, consisting of 344 exposures. The MUSE observations of the MUDF have been fully completed now. In this work, we use all the 358 MUSE exposures ($\approx$143\,h in total) of the MUDF survey. The final sample consists of 533 galaxies in MAGG, and 91 galaxies in MUDF, or a total of 624 galaxies. The median redshift of the sample is 1.0, and the median stellar mass is $10^9$\,\msun. The typical uncertainty in the redshift estimate is $\approx$60\,\kms, and the uncertainty in the base 10 logarithm of the stellar mass is typically between 0.1 and 0.2 dex.

The physical properties of the galaxies such as stellar mass and star formation rate (SFR) are obtained from jointly fitting the MUSE spectroscopy and photometry with stellar population synthesis (SPS) models using the Monte Carlo Spectro-Photometric Fitter \citep[MC-SPF;][]{Fossati2018}. In the case of the MUDF galaxies, we also use the optical and near-infrared photometry from the Hubble Space Telescope (HST) five band imaging \citep{Revalski2023}, and recently acquired K-band photometry from HAWK-I observations to constrain the spectral energy distribution (SED) fits. We refer to \citet{Fossati2019}, \citet{Dutta2020}, and \citet{dutta23} for a detailed description of the SED fitting procedure.

\subsection{\mgii Spectra}
\label{sec:observations_mgii}

To obtain the stacked \mgii spectra, we stacked the MUSE cubes in four different stellar mass bins (see Table~\ref{tab:stack_sample}) following the procedure explained in \citet{dutta23}. To compare with the models, we extracted integrated 1D spectra from the median stacked cubes, within a circular aperture of radius 10\,kpc, which we term as the core spectra, and within a circular annuli between radius 10 and 30\,kpc, which we term as the halo spectra. The $1\sigma$ errors on the flux are obtained from a bootstrapping analysis where we repeat the stacking 100 times with repetition. The above choices of apertures for spectral extraction are based on the typical extent of the continuum and \mgii\ line emission observed in the stacked narrow-band images and surface brightness profiles in our previous study \citep{dutta23}.

\begin{table}
	\caption{Galaxy sample used for stacking \mgii\ in different stellar mass bins.}
	\label{tab:stack_sample}
	\begin{tabular}{ccccc} 
		\hline
		  \mstar\ bin (log; \msun) & Number & Median $z$ & Median \mstar\ (log; \msun) \\
		\hline
        7--8   & 52  & 0.8 & 7.7  \\
        8--9   & 190 & 1.0 & 8.6  \\
        9--10  & 240 & 1.2 & 9.4  \\
        10--12 & 137 & 1.0 & 10.4 \\
		\hline
	\end{tabular}
\end{table}

In addition to the stacked spectra, we also study the \mgii\ line in the spectra of individual galaxies. To identify the \mgii\ line, we consider the galaxies with continuum S/N $\ge2$ in the MUSE integrated 1D spectra. The continuum S/N is estimated as the median S/N per pixel (1.25\,\AA) over two 10\,\AA\ windows on either side of the \mgii\ doublet lines. After removing the spectra that show broad \mgii\ emission typical of active galaxies, and those that are affected by strong sky line subtraction residuals at the \mgii\ wavelengths, we have a sample of 245 galaxies in MAGG and 70 galaxies in MUDF with continuum S/N $\ge2$. We detect \mgii\ in 168 of the galaxies (see Table~\ref{tab:mgii_class}).

To classify the \mgii\ profiles as either absorbers, emitters, or P Cyngni-like (redshifted emission and blueshifted absorption), we utilize a combination of both equivalent width cuts and visual inspection (by RD). To estimate the equivalent widths, we first convert the spectrum to rest-frame \mgii\ velocity. We fit a straight line to the spectral region within $\pm$3000\,\kms\ but outside of $\pm$1500\,\kms\ of the systemic velocity after applying 3$\sigma$ median clipping. Using this fit as the local continuum, we obtain the normalized spectrum. The equivalent width of the first \mgii\ line, $\lambda$2796, is estimated from $-500$\,\kms\ to the mid-point of separation between the doublet lines (385\,\kms). The equivalent width of the second \mgii\ line, $\lambda$2803, is estimated from the mid-point of doublet separation to 500\,\kms\ redward of the second line (1270\,\kms). Error in equivalent width is estimated by propagating the error in flux of each pixel. We adopt the following classification scheme for the \mgii\ profiles: \newline
(1) Absorber: Visual inspection plus both the \mgii\ doublet lines are detected with positive equivalent width at $\ge2\sigma$ significance; \newline
(2) Emitter: Visual inspection plus both the \mgii\ doublet lines are detected with negative equivalent width at $\ge2\sigma$ significance; \newline
(3) P Cygni: Visual inspection, redshifted emission lines, and absorption in at least one of the \mgii\ doublet lines; \newline
(4) Non-detection: \mgii\ profiles that do not satisfy the above conditions.

\begin{figure*}
    \centering
    \includegraphics[width=0.49\textwidth]{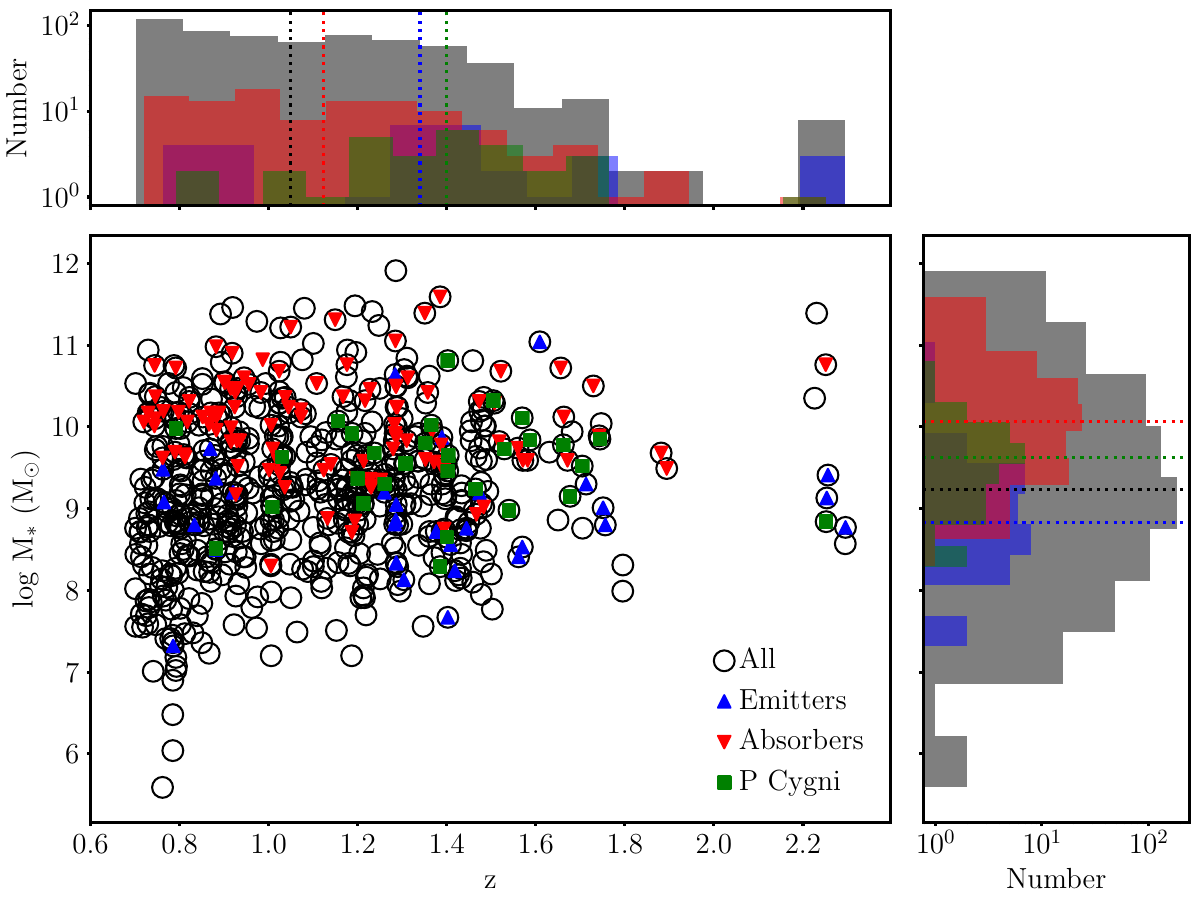}
    \includegraphics[width=0.49\textwidth]{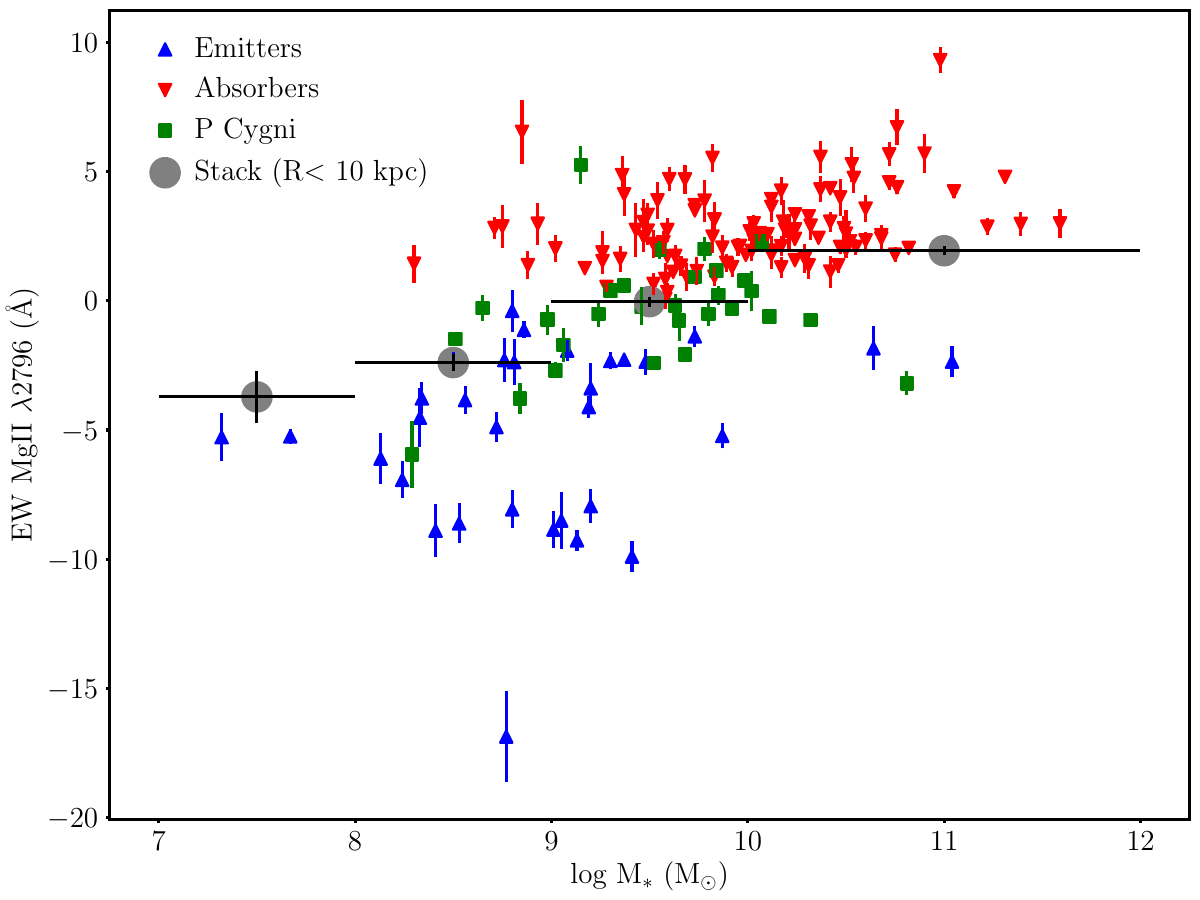}
    \caption{{\it Left:} The stellar mass as a function of redshift for all the galaxies (open black circle) in MAGG and MUDF with \mgii\ coverage in the redshift range, $0.7 \le z \le 2.3$. The galaxies with \mgii\ detection in the 1D spectra are marked by upward blue triangles for Emitters, downward red triangles for Absorbers, and green squares for P Cygni profiles. The histograms of the stellar mass and redshift distributions are shown in the right and top panels, respectively. The dotted line marks the median property of the sample. The color scheme of the histograms and dotted lines is the same as that of the symbols.
    {\it Right:} The equivalent width of the \mgii\ $\lambda$2796 line as a function of the stellar mass for the Emitters, Absorbers, and P Cygni sub-samples, following the same color scheme as in the left panel. The equivalent width estimates obtained from stacking the spectra of all galaxies (shown by open black circles in the left panel) in four different stellar mass bins, extracted within a central circular aperture of radius 10\,kpc, are also shown in grey circles.
    }
    \label{fig:obs_prop}
\end{figure*}

The properties of the galaxy sample used for \mgii\ stacking and of the galaxies with individual \mgii\ detections are summarised in Figure~\ref{fig:obs_prop}. From Table~\ref{tab:mgii_class} and the left panel of Figure~\ref{fig:obs_prop}, it can be seen that galaxies showing \mgii\ in absorption have higher stellar mass on average compared to those showing \mgii\ in emission. In contrast, the galaxies with P Cygni-like \mgii\ profiles have average stellar mass in between the two. The equivalent width of \mgii\ shows an increasing trend with the stellar mass, as seen from the right panel of Figure~\ref{fig:obs_prop}. This is consistent with the average trend obtained from the stacked samples in different stellar mass bins. Similar results have been found for the sample of galaxies with \mgii\ detection in the MUSE Hubble Ultra Deep Field Survey \citep{Feltre2018}.

In this work, we investigate the physical conditions and mechanisms in different galaxy populations that could be giving rise to the observed results through radiative transfer modeling of both the stacked and individual \mgii\ spectra. 
In the following section, we introduce our modeling and the variations of \mgii spectra for various parameters.

\begin{table}
	\caption{Classification of \mgii\ profiles in individual galaxy spectra}
	\label{tab:mgii_class}
	\begin{tabular}{cccc} 
		\hline
		  Sample & Absorbers & Emitters & P Cygni \\
		\hline
		MAGG & 92  & 12 & 18 \\
		MUDF & 15  & 20 & 11 \\
		  Both & 107 & 32 & 29 \\
        \hline
        Median $z$ & 1.1 & 1.3 & 1.4 \\
        Median \mstar\ (log; \msun) & 10.0 & 8.8 & 9.6 \\
        \hline
	\end{tabular}
\end{table}

\section{Radiative Transfer Modeling for \mgii spectra}
\label{sec:modeling}

\subsection{Spectral fitting pipeline}

\begin{table*}
\caption{Free parameters used in the \mgii radiative transfer simulations. The total number of simulated spectra is 201,600.}
\centering
\begin{tabular}{llll}
\hline
         Parameter   & Range      & Bins                                                & Note \\ \hline
 \Nmg        & $10^{12.5} - 10^{17} \unitNHI$& $0.5\,$dex               & \mgii column density \\
 \vexp       &  $-200-500 \kms$& $50 \kms$              & expansion velocity \\
 \sigran     &  $25-200 \kms$& $25 \kms$              & random speed of \mgii medium \\
 \sigsrc     &  $25-200 \kms$& $25 \kms$              & width of intrinsic \mgii emission  \\
 \EWint     &  $0-20$ \AA& 1 \AA\              & equivalent width of intrinsic \mgii emission  \\
\hline
\end{tabular}
\label{tab:par_sim}
\end{table*}

\begin{figure*}
    \centering
    \includegraphics[width=\textwidth]{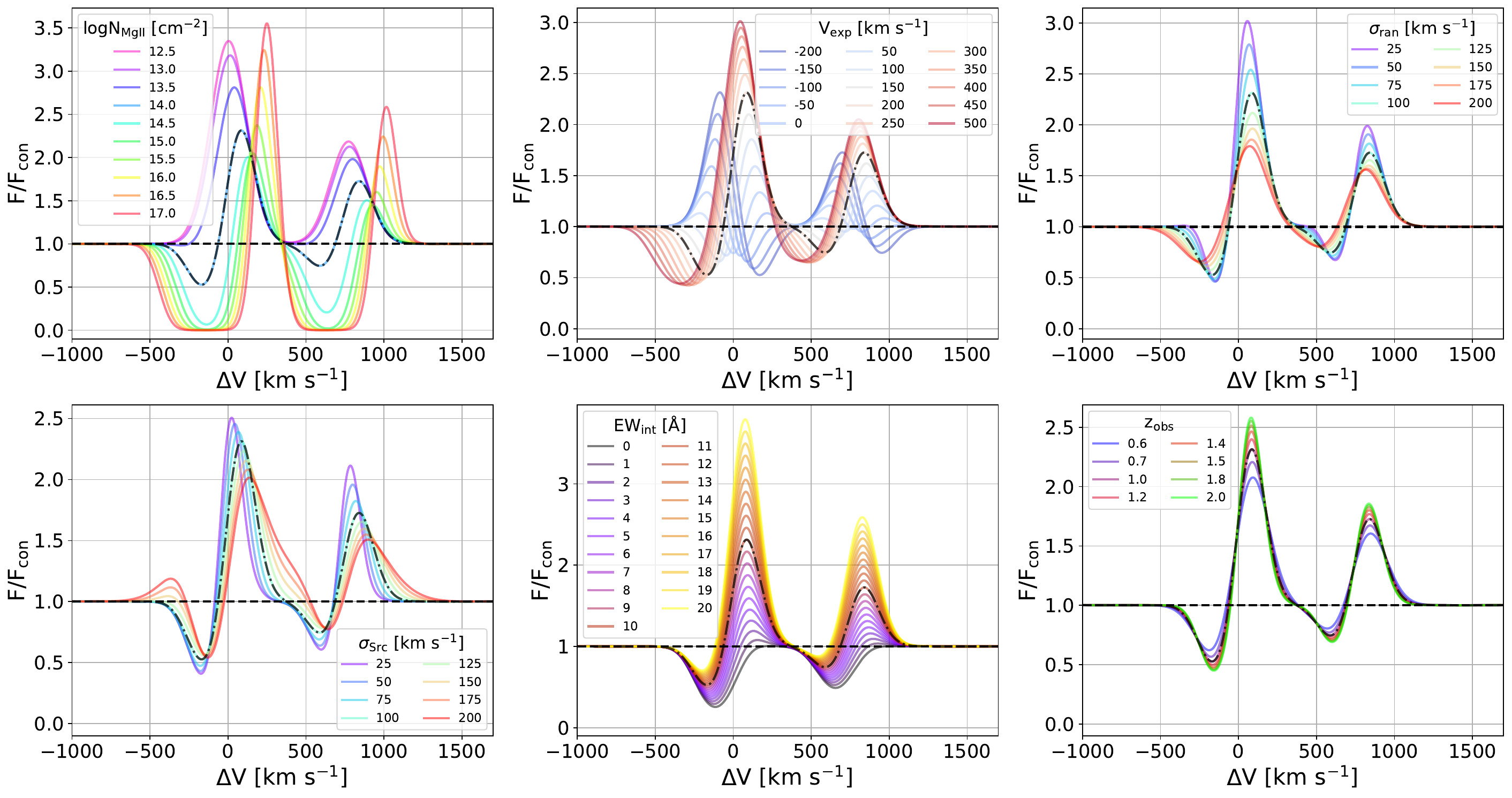}
    \caption{
    Simulated \mgii doublet spectra as a function of the Doppler factor $\Delta$V from the line center of the \mgii K line using the 3D Monte-Carlo radiative transfer simulation \texttt{RT-scat}. The black dashed, dotted line is the fiducial spectrum at \Nmg = $10^{14} \unitNHI$, \vexp = 200 \kms, \sigran = 100 \kms, \sigsrc = 100 \kms, \EWint  = 10 \AA\, and \zobs = 1.
    Each panel showcases the spectral behavior varying six parameters, \Nmg, \vexp, \EWint, \sigran, \sigsrc, and \zobs. 
    }\label{fig:spec_par}
\end{figure*}

To analyze the observed \mgii signal, we utilize the simulated spectra generated by the 3D Monte-Carlo simulation \texttt{RT-scat} \citep{chang23a,chang24}. 
Specifically, we adopt a spherical model composed of a spherical \mgii halo with varying expanding velocity and a point source at the center of this spherical halo. 
We assume two types of intrinsic spectral distributions near the \mgii doublet: a flat continuum, representing the stellar continuum near 2800~\AA, and Gaussian-like emission with a fixed flux ratio of \mgii $\lambda$2796 (K line) \& $\lambda$2803 (H line) at 2 and width of $\sigma_{\rm Src}$.
The outer radius of the halo, denoted as $R_{\rm H}$, is fixed at 100 kpc; the inner radius is 1 kpc, corresponding to $0.01 R_{\rm H}$.
This choice is made arbitrarily, but as the radiative transfer is affected by the intercepting column densities (and kinematics) and not by the physical distances, the solutions found here are generally valid and can be rescaled to any other halo radius.
The radial velocity of the outflow is proportional to the radius from the central source, i.e., $v(r) = \vexp r/R_{\rm H}$.
We generate 201,600 \mgii spectra across a parameter range outlined in Table~\ref{tab:par_sim}, considering five parameters:
\begin{itemize}
\item The \mgii column density (\Nmg) measured from the center to $R_{\rm H}$. Specifically, we fill the spherical cold gas halo from the inner radius $0.01 R_{\rm H}$ to the outer radius $R_{\rm H}$ with a constant \mgii number density $n_{\mgii} = \Nmg/0.99R_{\rm H}$.

\item The maximum outflow velocity (\vexp) reached the edge of the halo.

\item The random motion of \mgii parametrized by \sigran. Note that we employ the `microturbulent approximation' in this work (as done in many previous studies; e.g., \citealp{verhamm06, Gronke15,seon20}), and thus a given random velocity corresponds to an effective temperature $T_{\rm eff}\approx 3000\sigma_{\rm ran}^2\,{\rm K\,s}^2{\rm km}^{-2} + 10^4\,$K.

\item The two parameters describing the intrinsic emission from the point source, namely the width of the intrinsic \mgii line (\sigsrc) and the equivalent width of intrinsic emission (\EWint).
\end{itemize}

In order to compare the observed and simulated spectra, we have to take into account that the emergent spectrum is spatially varying. To mimic the observations closely, we use the `core' spectrum, that is, the observed spectrum within $10\,$kpc projected distance as our fiducial integrated spectrum and refer to it henceforward as the `simulated' spectrum. When modeling the full spatially varying spectrum, we define which range of projected radii are being used.
In addition, we adopt the spatial resolution of MUSE ($\approx 6$ kpc at $z \approx 1$) in the simulated spectra through 2D Gaussian convolution and the spectral resolution of MUSE as a function of wavelength.

To obtain the best fitting set of the five parameters listed in Table~\ref{tab:par_sim}, we compare the observed and simulated spectra using a chi-square test in wavelength space.
The details of the fitting method are described in Appendix~\ref{sec:fitting}.

\subsection{Impact of the individual parameters}
In this section, we highlight the change in the individual parameters described above on the emergent spectrum. While this allows for a better understanding and interpretation of the modeling results, we want to note that degeneracies and additional dependencies do exist (which are captured by our fitting pipeline). A more detailed exploration of the parameter space as well as general \mgii radiative transfer is presented in \citet{chang24}.

Figure~\ref{fig:spec_par} shows the dependencies of the \mgii spectra on various parameters.
In the left top panel, as the \mgii column density (\Nmg) increases, the absorption feature gets stronger.
Particularly, the variation of a spectral profile is large at \Nmg = $10^{13.5 - 14.5} \unitNHI$, where the optical depth of \mgii at the line center increases from $\sim 1$ to $\sim 10$. The regime for this large variation depends on other parameters, namely \sigr and \vexp, which affect the optical depth of the \mgii doublet. Higher \sigr and \vexp enhance the \Nmg range in $\Nmg = 10^{14-15} \unitNHI$.

In the central top panel of Figure~\ref{fig:spec_par}, the expansion velocity (\vexp) induces the broadening of the absorption feature due to the distribution of the radial velocity. Since the radial velocity is proportional to the distance from the central source, a higher \vexp causes a wider absorption in the velocity range from 0 \kms to $-$\vexp. 
The spectra with a positive and negative \vexp, representing outflow and inflow, show a blue- and red-shifted absorption feature, respectively.

Concerning the random motion (\sigr) and intrinsic emission width (\sigsrc) in the right top and left bottom panels of Figure~\ref{fig:spec_par}, respectively, their influences are comparatively similar. Increasing these parameters results in a broader red peak since a higher \sigsrc intrinsically enhances the width of emission features, and \sigr also causes the line broadening via scattering.
Increasing the equivalent width of intrinsic emission (\EWint) strengthens the emission features.
In the bottom right panel of Figure~\ref{fig:spec_par}, we examine the dependence on observed redshift \zobs. 
As the MUSE spectral resolution increases with increasing wavelength, the spectrum for a higher \zobs is slightly narrower than that for a lower \zobs.\\

In summary, while our modeling approach varies all parameters simultaneously and thus takes the full spectral shape into account, the most constraining parts are the absorption features for the scattering medium (i.e., \Nmg, \vexp and \sigr) and the emission part of the observed \mgii spectra for the emission parameters ($\sigma_{\rm Src}$ and \EWint).
In the following section, we interpret the stacked \mgii data based on the spectral behaviors illustrated in Figure~\ref{fig:spec_par}.

\begin{figure}
    \centering
    \includegraphics[width=\columnwidth]{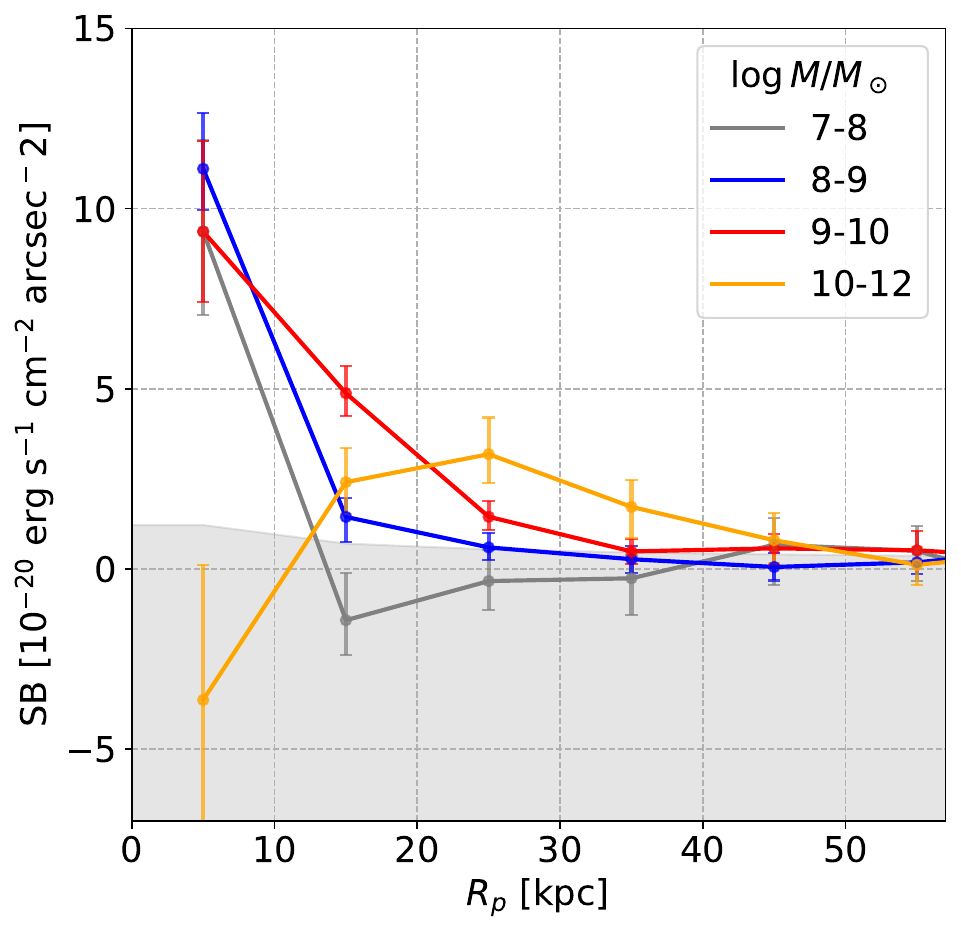}
    \caption{Azimuthally averaged surface brightness of the stacked \mgii emission as a function of projected radius \rp.
    The line colors represent the range of stellar mass bin of the stacked data, $\log \mstar/\msun $ = $7-8$ (green), $8-9$ (blue), $9-10$ (red), and $10-12$ (orange).
    The grey zone is the $3\sigma$ detection limit of the surface brightness as a function of \rp.
    }
    \label{fig:stack_SB}
\end{figure}

\begin{figure*}
    \centering
    \includegraphics[width=\textwidth]{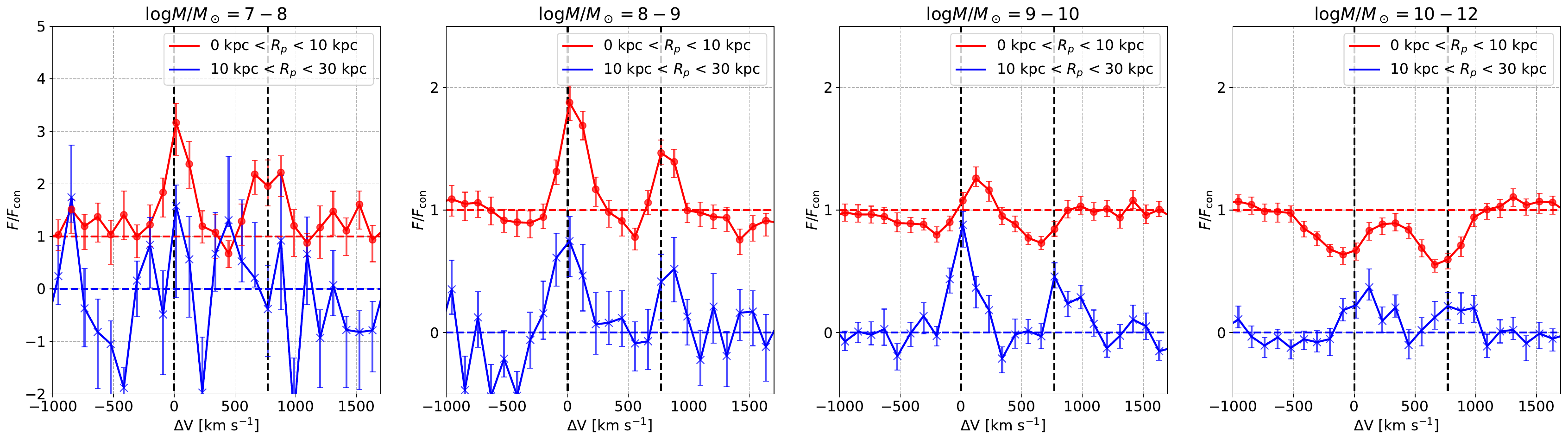}
    \caption{Stacked core spectrum in $\rp < 10\, \rm kpc$ (red), and stacked halo spectrum in \rp from 10 kpc to 30 kpc (blue), as a function Doppler factor $\Delta V$ of the \mgii K line.
    Each panel represents different stellar mass bins, $\log \mstar/\msun $ = $7-8$, $8-9$, $9-10$, and $10-12$.
    The stacking halo spectra are continuum subtracted. 
    The core and halo spectra are normalized by dividing by the continuum of the core spectrum for each mass bin.
    The red and blue dashed horizontal lines represent the normalized continuum and the subtracted continuum (i.e., $\sim 0 $ \ergscmarc) near the \mgii doublet, respectively.
    The black vertical lines represent the line center of the \mgii K and H lines.
    }
    \label{fig:stack_spec}
\end{figure*}

\section{Dependence of Stacked \mgii on Stellar Mass}
\label{sec:stacking_observation}

In this section, we explore the dependence of the stacked \mgii data on stellar mass.
Figure~\ref{fig:stack_SB} presents radial profiles of surface brightness (SB) for the \mgii doublet for four distinct $\log \mstar/\msun$ bins: 7-8, 8-9, 9-10, and 10-12. The SB values are the azimuthally averaged values in circular annuli of radius 10 kpc from the galaxy centre.
Here, negative and positive values of SB denote \mgii absorption and emission features, respectively. The SB profiles are obtained from narrow-band images of the stacked \mgii emission over the velocity range 0 to 300 \kms around both the \mgii K and H line centroids, after subtracting the continuum emission. 
As shown in Figure~\ref{fig:obs_prop}, when \mstar increases, the spectrum shows stronger absorption features.
Thus, in the core region at $R_p < 10$ kpc, the SB for \mstar/\msun $=10^{10-12}$ is negative due to a strong absorption.
However, in the halo region at $R_p > 20\, \rm kpc$, the trend differs from that in the core region. In Figure~\ref{fig:stack_SB}, the SB profile becomes more extended with increasing \mstar.
This is because \mgii emission is spatially extended to the halo through scattering with \mgii atoms in the cold medium surrounding a galaxy.

\subsection{\mgii Spectra in Core and Halo Regions}
\label{sec:obs_profile}

Figure~\ref{fig:stack_spec} shows the core (projected radius, $R_p$ = 0-10 kpc) and halo ($R_p$ = 10-30 kpc) spectra 
for different stellar mass bins. Here, the $y$-axis is the observed flux divided by the continuum, $F/F_{\rm con}$.
As also shown in Figure~\ref{fig:stack_SB}, the strength of the absorption feature increases with increasing \mstar.
The core spectrum at \mstar/\msun $= 10^{7-8}$ exhibits clear emission features, while the P-Cygni profile starts to appear at \mstar/\msun $=10^{8-9}$ and is prominent at \mstar/\msun $=10^{9-10}$. Notably, strong \mgii absorption without any emission feature is evident at \mstar/\msun $=10^{10-12}$.
Therefore, this trend shows that the rich cold gas within and around a massive galaxy causes a strong absorption feature of the \mgii spectra within the galaxy.

The halo spectrum is intricately linked to the absorption in the core spectrum.
In the left panel of Figure~\ref{fig:stack_spec} for \mstar/\msun $\ge 10^{8}$, where core spectra exhibit significant absorption features,
\mgii emission becomes distinctly detectable in the halo. 
In the low \mstar regime ($< 10^8 \msun$), the absorption (emission) feature of \mgii is negligible in the core (halo) spectrum. 
This indicates a clear correlation between absorption in the core region and emission in the halo.
For instance, in the case of optically thin cold gas for \mgii (\Nmg $< 10^{13} \unitNHI$), \mgii scattering effects are weak, resulting in an observed \mgii emission profile without absorption features and a negligible emission in the halo.
Conversely, in cases of optically thick cold gas for \mgii (\Nmg > $10^{14} \unitNHI$) surrounding galaxies, the gas in the observed direction and along its perpendicular direction causes absorption features in the core spectrum and emission features in the halo spectrum, respectively.
Thus, abundant cold gas induces absorption in the core region and emission in the halo.

However, in the right panels of Figure~\ref{fig:stack_spec}, the halo spectrum at \mstar/\msun $= 10^{9-10}$ is stronger than that at \mstar/\msun $= 10^{10-12}$, although its absorption is weaker than that for the higher stellar mass. We expect that this originates from the dusty cold medium, angular distribution of cold gas around massive galaxies, or \mgii emission spatially extended to a larger radius.
We will investigate the weakening of the halo spectrum at the highest mass regime later in \S~\ref{sec:modeling_halo} through the fitting results of the stacked spectra.


\subsection{Characterization of \mgii K and H Line Absorption}
\label{sec:obs_absorption}

In general, when both the K and H lines at 2796 and 2803 \AA, are not optically thick, the absorption of the K line is typically stronger than that of the H line due to the scattering cross section of the K line being twice as high. In a medium that is highly optically thick for both lines ($\tau_0 \gg 1$), the absorption features become saturated, and both K and H lines exhibit similar absorption profiles.
Thus, the K and H lines' absorption strength ratio is always between 2 to 1.

However, in the right panels of Figure~\ref{fig:stack_spec}, the core spectra for $\mstar/\msun = 10^{9-10}$ and $10^{10-12}$ show stronger absorption in the H line than in the K line. This serves as evidence of significant intrinsic \mgii emission.
If the \mgii emission arises from recombination or collisional excitation, the \mgii K line emission should be approximately twice as luminous as the H line emission. The brighter \mgii K emission, when mixed with saturated absorption due to low spectral resolution or stacking process, can result in weaker absorption in the observed K line compared to the H line.
For this reason, we expect that massive (\mstar/\msun $> 10^9$) galaxies exhibit both a strong stellar continuum and intrinsic \mgii emission.
We will discuss further details through our modeling in the following section.

\section{Results}
\label{sec:result}

\begin{figure*}
    \centering
    \includegraphics[width=\textwidth]{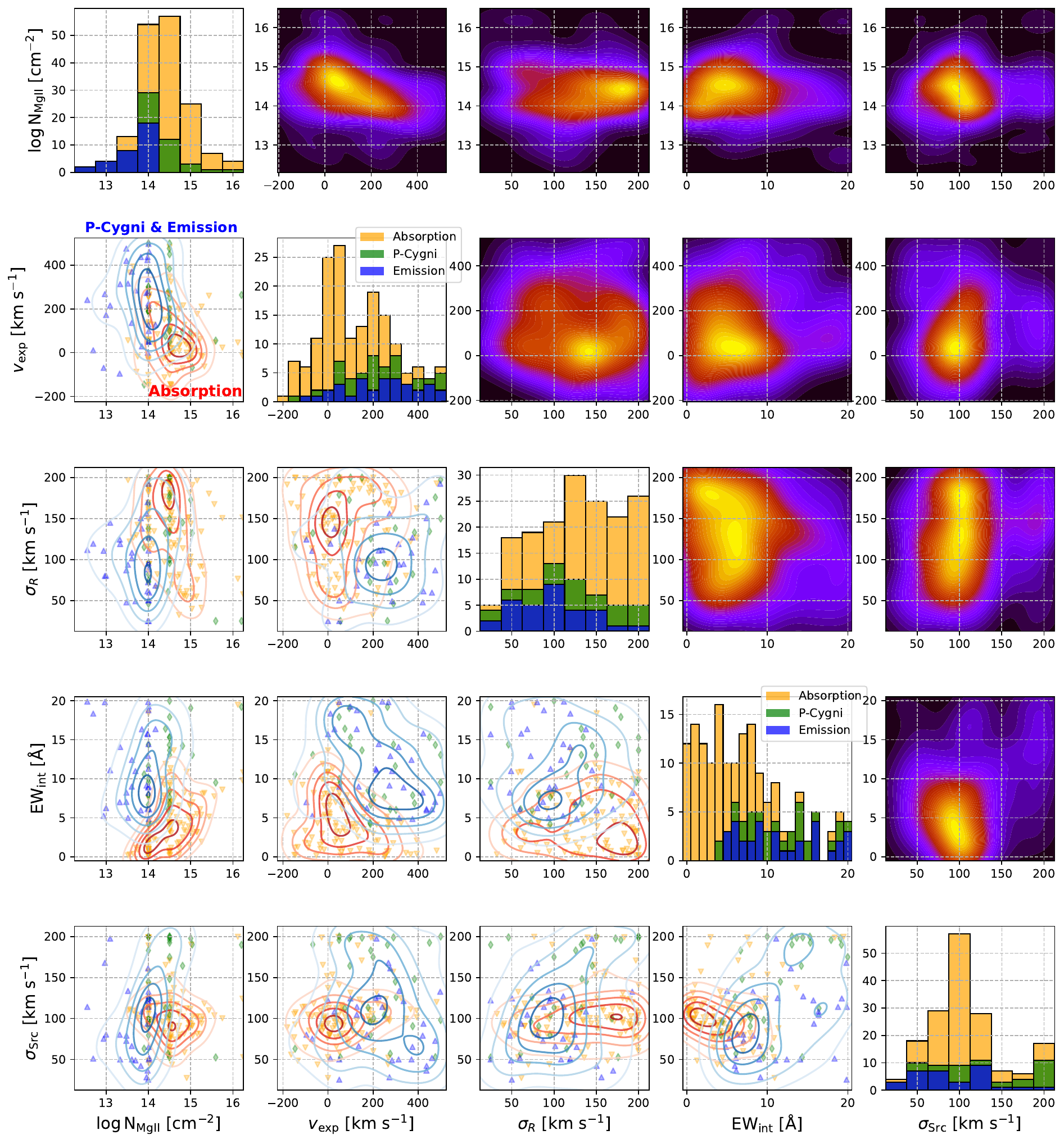}
    \caption{Histograms of fitting parameters, \Nmg, \vexp, \sigr, \EWint, and \sigsrc.
    The panels along the diagonal show the stacked histograms of each parameter.
    The colors represent the type of spectrum: `Absorption' (orange), `P-Cygni' (green), and `Emission' (blue).
    The panels in the lower triangle show the density map correlation between the two parameters.  }
    \label{fig:2d_map}
\end{figure*}

This section analyzes the fitting results of the observed spectra, including individual detections and stacks.
Figure~\ref{fig:2d_map} shows the histogram of and correlations between five fitting parameters: \Nmg, \vexp, \sigr, \EWint, and \sigsrc. We explore the distribution of fitting parameters and their correlation through the fitting results of individual detections. Note that our pipeline manages to reproduce observed spectra very well. Examples of fitting individual detections, as well as a discussion of the fit quality, are presented in Appendix~\ref{sec:fitting}.

From the panels in the upper triangle of Figure~\ref{fig:2d_map}, there is a strong correlation only between the parameters, \Nmg and \vexp.
On the other hand, in the panels in the lower triangle of Figure~\ref{fig:2d_map}, the correlations of the absorption spectra (red) and P-Cygni \& emission spectra (blue) are separately distributed.
This indicates that the type of the spectrum is crucial to understanding the physical mechanism giving rise to the \mgii spectrum.
Therefore, we compare the distributions of parameters for each spectral type: `Absorption', `P-Cygni', and `Emission' as defined in \S~\ref{sec:observations}.

\subsection{Analysis of Fitting Parameter Trends}\label{sec:fitting_result}

The \mgii column density, \Nmg, is a prominent indicator of the strength of an absorption feature.
The spectra at $\Nmg < 10^{14} \unitNHI$ have a negligible absorption feature as can be seen in the left top panel of Figure~\ref{fig:spec_par}.
In the first top panel of Figure~\ref{fig:2d_map}, the distributions from `absorption' (orange), `P-Cygni' (green), and `emission' (blue) spectra indicate that the behavior of simulated spectra for various \Nmg is well reflected. 
For the absorption and P-Cygni spectra where an absorption feature appears, \Nmg of the absorption cases cluster around $10^{14.5} \unitNHI$ and \Nmg for P-Cygni cases $ \ge 10^{14} \unitNHI$.
\Nmg of the emission spectra is skewed distribution toward a lower value in the vicinity of $\Nmg \le 10^{14} \unitNHI$.
In addition, our modeling indicates that half of the emission cases have a negligible absorption feature at \Nmg < $10^{14} \unitNHI$.
Consequently, \Nmg of absorption spectra is higher than that of emission spectra.

The expansion velocity, \vexp, is estimated by the location of the dip in the \mgii absorption profile and represents the kinematics of optically thick cold gas (\Nmg $\ge 10^{14} \unitNHI$).
In the second panel on the diagonal from the top-left to the bottom-right of Figure~\ref{fig:2d_map}, the histogram of \vexp has two peaks at 0 \kms and 200 \kms.
This indicates that \mgii absorption originates from two types of cold gas: static, slow cold gas (\vexp $< 100$ \kms) and outflowing galactic wind (\vexp $> 200$ \kms).
The absorption spectra dominate the primary peak at 0 \kms and the inflow regime (i.e., negative \vexp). 
Most galaxies with a high stellar mass (\mstar/\msun $> 10^{10}$) show \mgii in absorption, as shown in Figure~\ref{fig:obs_prop}.
Thus, the abundant static cold ISM or infalling gas of massive galaxies causes the absorption spectrum without an emission feature at $\rp~<~10$~kpc.

The random motion of cold gas, \sigr, represents the width of the absorption profile.
In the third panel on the diagonal of Figure~\ref{fig:2d_map},
\sigr of P-Cygni and emission spectra are broadly distributed from 0 to 200 \kms. 
In the absorption case, \sigr is clustered towards higher values.
These distributions indicate that static, slow cold gas (absorption spectra) has faster random motion than the outflowing gas (P-Cygni and emission spectra). But, in the second panel on the third row of Figure~\ref{fig:2d_map}, P-Cygni and emission cases (blue triangles) show a positive correlation between \vexp and \sigr.

The equivalent width of intrinsic \mgii emission, \EWint, represents the strength of the \mgii emission from the source.
In the fourth panel on the diagonal of Figure~\ref{fig:2d_map},
\EWint of the absorption spectra are broadly distributed at \EWint $< 10$ \AA. 
This is because the absorption-only spectrum can also indicate the presence of intrinsic emission, which is washed out by the absorption. Thus, the intrinsic emission can be estimated by comparing the absorption line ratio of K and H lines as discussed in \S~\ref{sec:obs_absorption}.
Furthermore, the same panel shows that \EWint of the P-Cygni and emission spectra are higher than 5 \AA\, to reproduce their emission features.

The fifth panel on the diagonal of Figure~\ref{fig:2d_map} shows that the width of the intrinsic emission, \sigsrc, of absorption spectra is clustered at 100 \kms. 
However, measuring \sigsrc from the absorption spectra is challenging because of the absence of a clear emission feature in the whole spectrum. Thus, it is hard to interpret the physical properties of the observed spectra from the \sigsrc of the absorption spectra.
\sigsrc of P-Cygni spectra is higher than that of the emission spectra since a higher \sigsrc enhances the red spectral peak, such as in outflowing gas, as shown in the left bottom panel of Figure~\ref{fig:spec_par}.

\begin{figure*}
    \centering
    \includegraphics[width=\textwidth]{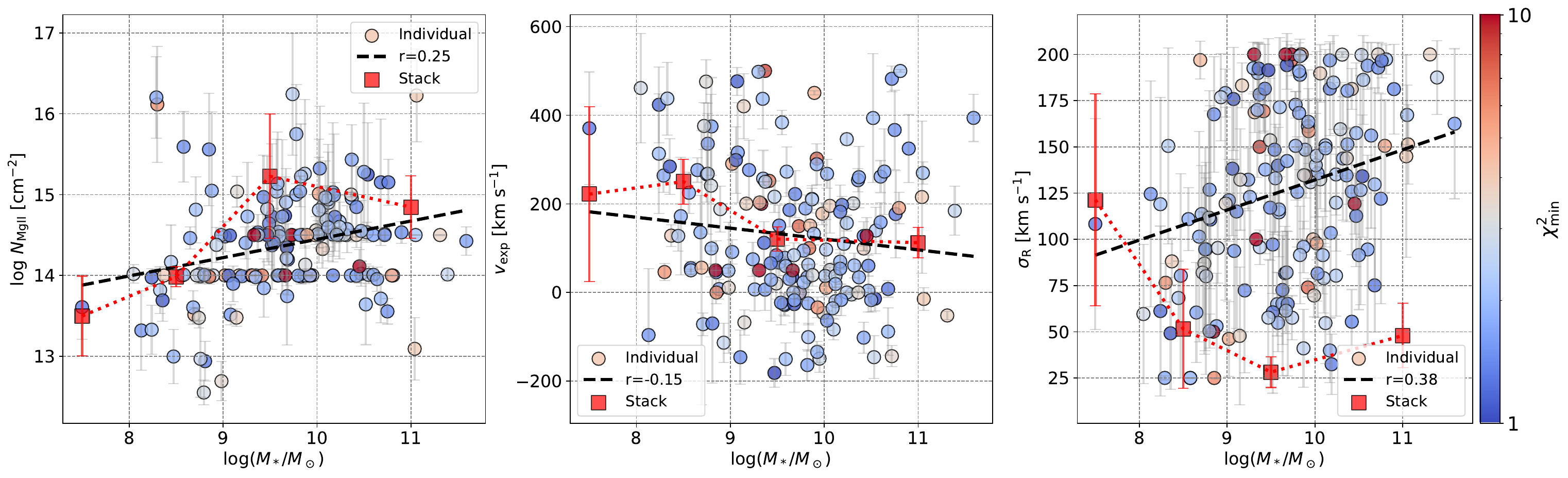}
    \caption{Fitting parameters characterizing the cold gas as a function of stellar mass \mstar: \mgii column density \Nmg (left), expansion velocity \vexp (center), and the random motion of cold gas \sigr (right). The circle colors indicate the minimum $\chi^2$ value, with error bars showing $\pm$ weighted standard deviation (see \S~\ref{sec:fitting}). Red squares represent the fitting results for stacked spectra across four stellar mass bins: $\log M_* / M_\odot = 7-8$, $8-9$, $9-10$, and $10-12$.
    The inclination of the black dashed lines represents the linear correlation between fitting parameters and $\log \mstar/\msun$. Pearson correlation coefficient $p$ for $\log \Nmg$, \vexp, and \sigr are 0.25, $-0.15$, and 0.38, respectively.
    }
    \label{fig:trend_mass}
\end{figure*}

From the above results, it is clear that the distributions of the five parameters depend on the type of the observed \mgii spectrum.
Particularly, the distribution parameters of the absorption spectra are different from those of P-Cygni \& emission spectra.
The panels in the lower triangle of Figure~\ref{fig:2d_map} show separately distributed density maps for the correlations between two parameters of the absorption spectrum (red) and P-Cygni \& emission spectra (blue).
Compared with the P-Cygni \& emission cases, the absorption cases have higher \Nmg and \sigr, and lower \vexp and \EWint. 
This indicates that the absorption spectrum mainly originates from optically thick slow or inflowing cold gas with large random motion, and most P-Cygni and emission spectra come from the outflowing gas.
As most high stellar mass galaxies exhibit \mgii in absorption (see Figure~\ref{fig:obs_prop}),
we will explore the relation between the fitting parameters and stellar mass in the following section.

\subsection{Dependence of Fitting Parameters on Stellar Mass}

\subsubsection{Correlation between Stellar Mass and Cold Gas Properties}
\label{sec:cold_gas}

This section focuses on the correlation between the stellar mass, \mstar, and three critical parameters that characterize the cold gas: the \mgii column density (\Nmg), the expansion velocity (\vexp), and the random motion of the cold gas (\sigr).
Figure~\ref{fig:trend_mass} shows weight averages of \Nmg, \vexp, and \sigr, estimated from fitting with the simulated spectra as a function of stellar mass \mstar.
The left panel of Figure~\ref{fig:trend_mass} shows a positive correlation between \Nmg and \mstar with Pearson correlation coefficient $p = 0.25$.
In the low \mstar regime ($< 10^{9}\msun$), \Nmg informed from stacked spectra (red squares) and most individual detections (circles) are less than $10^{14} \unitNHI$. 
In the high \mstar regime ($> 10^{9}\msun$), \Nmg values are higher than $10^{14} \unitNHI$.  
This trend suggests that the absorption features in the stacked spectrum become stronger with increasing \mstar, as shown in Figure~\ref{fig:stack_spec}. 
Furthermore, most of the individual detections in the high \mstar regime are categorized by the absorption profile, as depicted in Figure~\ref{fig:obs_prop}. 

In the center panel of Figure~\ref{fig:trend_mass}, \vexp and \mstar show a negative correlation with $p = -0.15$.
\vexp in the low \mstar regime is largely scattered in \vexp from $-100$ \kms to 500 \kms.
Note that the measurements of \vexp in this low mass regime are uncertain and have a large error bar because most individual detections are classified as emission in Figure~\ref{fig:obs_prop}.
However, \vexp is predominantly distributed below $100 \kms$ in the high \mstar regime.
This pattern implies that such absorption could originate either from the cold ISM within galaxies or intrinsic \mgii absorption in the stellar continuum. 
Consequently, the different \vexp distributions in the high and low mass regimes indicate that the kinematics of the optically thick cold gas traced by the \mgii doublet depends on the stellar mass of the galaxies, as also shown by Figure~\ref{fig:2d_map}.

The right panel of Figure~\ref{fig:trend_mass} demonstrates that \sigr of individual detections increases with increasing \mstar. This trend is expected, as galaxies of higher stellar mass generally have more massive halos, which experience enhanced random motion of cold gas.
However, the measurements of \sigr have large error bars. 
It is challenging estimating \sigr accurately when the spectral resolution (spectral resolution of MUSE spectra $\approx$ 85-200 \kms at \zobs = 0.6 - 2.2, see Equation~\ref{eq:MUSE_resolution}) is higher than \sigr.
In addition, as the stacking process leads to additional broadening and effectively lowers the resolution, the measurements of \sigr of stacking spectra are also inaccurate.

\subsubsection{Intrinsic \mgii emission}
\label{sec:intrinsic_mgii}

\begin{figure*}
\begin{center}
\includegraphics[width=0.33\textwidth]{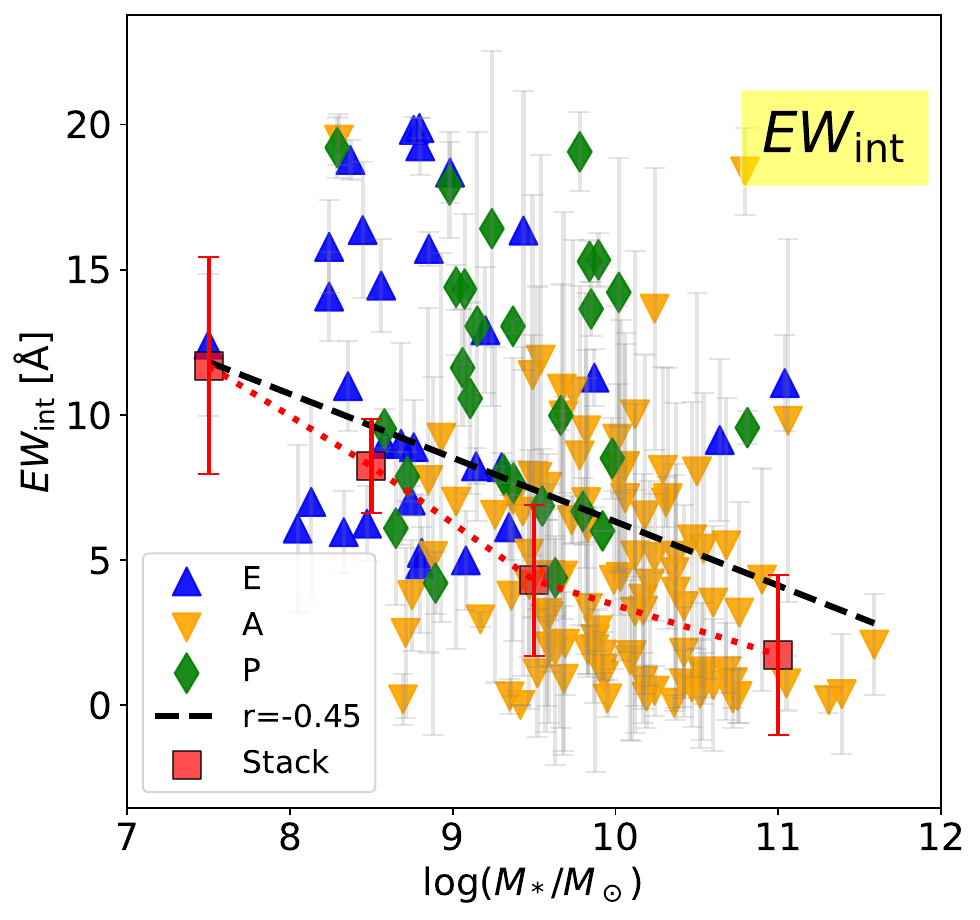}\includegraphics[width=0.33\textwidth]{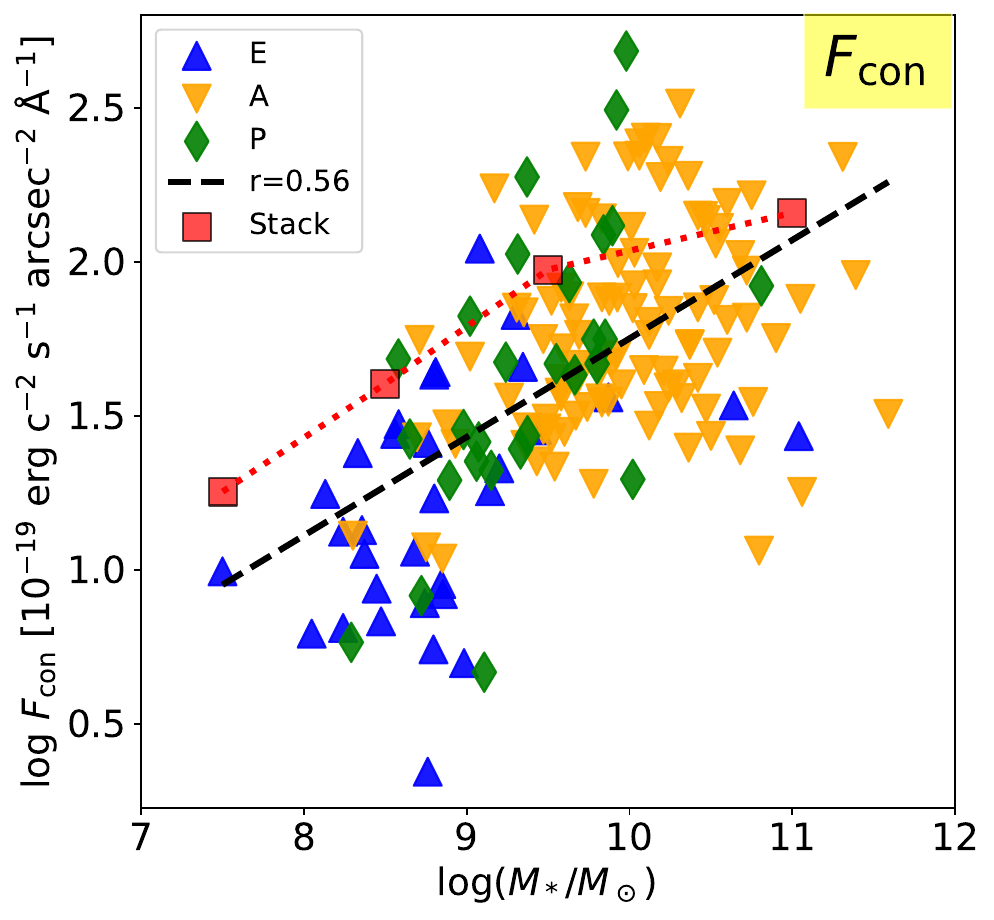}\includegraphics[width=0.33\textwidth]{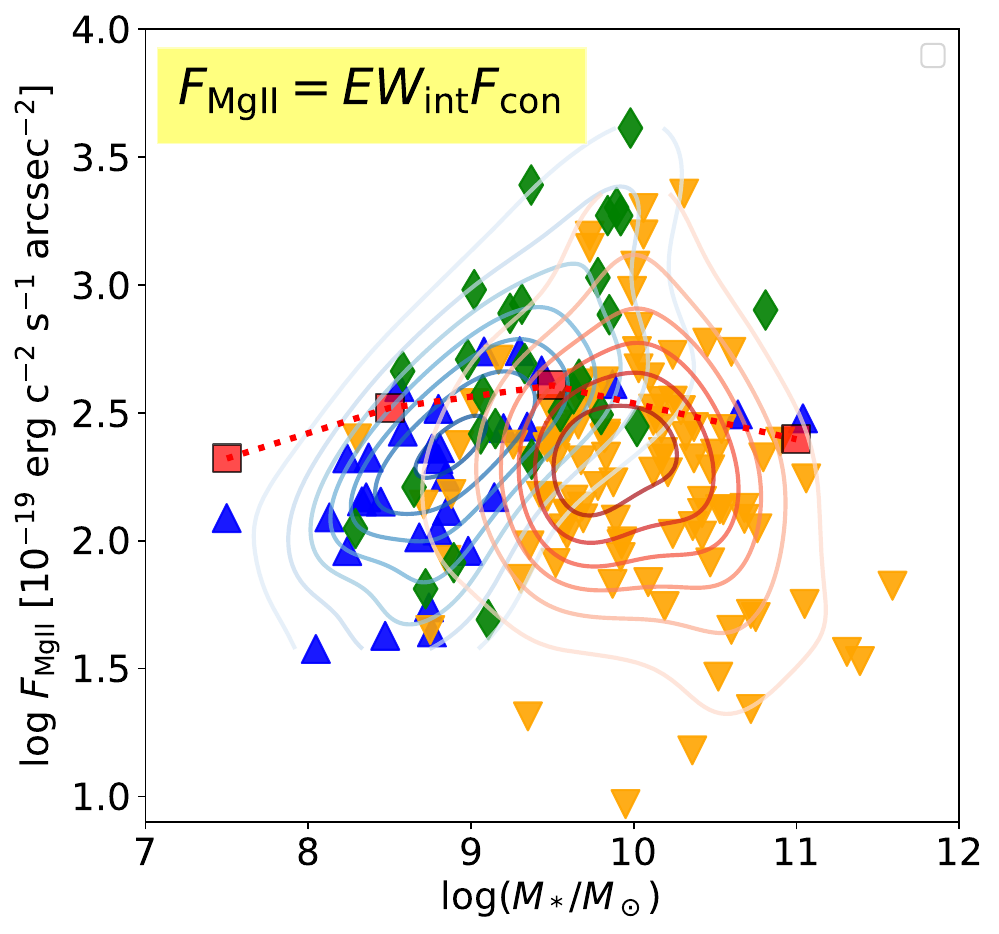}\\
\end{center}
\caption{The equivalent width of intrinsic \mgii emission \EWint (left), the continuum flux near \mgii emission $F_{\rm con}$ (center), and the flux of intrinsic \mgii emission $F_{\rm MgII}$ (right) as functions of \mstar. The blue triangles, orange nablas, and green diamonds represent individual spectra classified as 'Emission,' 'Absorption,' and 'P-Cygni,' respectively.
The inclination of the black dashed lines represents values of $p$, which are 0.45 and 0.56 for \EWint and $F_{\rm con}$, respectively.
}
\label{fig:EW_trend}
\end{figure*}


Due to its resonant nature, the formation of the \mgii line profile is determined by the properties of cold gas and intrinsic \mgii emission.
In the previous section, we explored the correlation between cold gas properties and the stellar mass of the galaxy.
This section focuses on the intrinsic \mgii emission parameters obtained through fitting of the emergent line shape.
The findings from \S~\ref{sec:obs_absorption} establish that spectra with K line (\mgii $\lambda2796$) absorption weaker than the H line (\mgii $\lambda2803$) distinctly suggest the existence of intrinsic \mgii emission. 
Two of the fitting parameters, the equivalent width and the width of the intrinsic \mgii emission, allow us to investigate the properties of the intrinsic emission.

Figure~\ref{fig:EW_trend} shows the fitting results for the equivalent width of intrinsic \mgii emission, \EWint, the stellar continuum near \mgii emission, $F_{\rm con}$, and the flux of intrinsic \mgii emission, $F_{\rm MgII}$, as a function of \mstar. 
In the left panel of Figure~\ref{fig:EW_trend}, \EWint of individual detections and stacked spectra shows a negative correlation (see the black dashed line). 
It is because \EWint of most individual cases in the high mass regime (\mstar $ > 10^{9} \msun$) are less than 5 \AA.
However, the trend of \EWint is insufficient to conclude that \mgii emission intrinsically decreases with increasing \mstar since, in general, more massive galaxies generate more \mgii photons.


To measure the total flux of the intrinsic \mgii emission, the continuum flux $F_{\rm con}$ is multiplied by \EWint.
The center panel of Figure~\ref{fig:EW_trend} shows that $F_{\rm con}$ increases as \mstar increases.
In the right panel of Figure~\ref{fig:EW_trend}, we calculate the intrinsic \mgii emission flux as $F_{\rm MgII}=\EWint F_{\rm con}$.
The $F_{\rm MgII}$ of individual detections are largely scattered across \mstar since \EWint and $F_{\rm con}$ show a positive and negative correlation with \mstar, respectively.
In addition, the trend of $F_{\rm MgII}$ with \mstar is not monotonic. The $F_{\rm MgII}$ of stacked spectra increases with increasing \mstar until \mstar/\msun $< 10^{10}$ and decreases at \mstar/\msun $= 10^{10-12}$.

To explore the trend of intrinsic \mgii emission $F_{\rm MgII}$, we separately checked the $F_{\rm MgII}$ distribution of 'absorption' and 'emission' \& 'P-Cygni' spectra.
The right panel of Figure~\ref{fig:EW_trend} shows the contour of the two distributions.
The 'emission' \& 'P-Cygni' detections positively correlate with \mstar (see the blue contours). The 'absorption' cases (the red contours) show a negative correlation with \mstar. 
The strength of intrinsic \mgii emission increases with increasing \mstar over $\mstar/\msun = 10^{7-9}$. However, it decreases at $\mstar/\msun > 10^{10}$ because abundant dusty cold gas within galaxies causes a weaker continuum and stronger intrinsic absorption.
In addition, we expect some of the large scatter in the distribution of $F_{\rm MgII}$ to originate from line-of-sight effects because the dust extinction of the radiation near \mgii emission along the edge-on direction of galaxies is higher than that along the face-on direction.

\subsection{Modeling Halo Spectra and Surface Brightness Profiles}\label{sec:modeling_halo}

\begin{figure*}
    \centering
    \includegraphics[width=\textwidth]{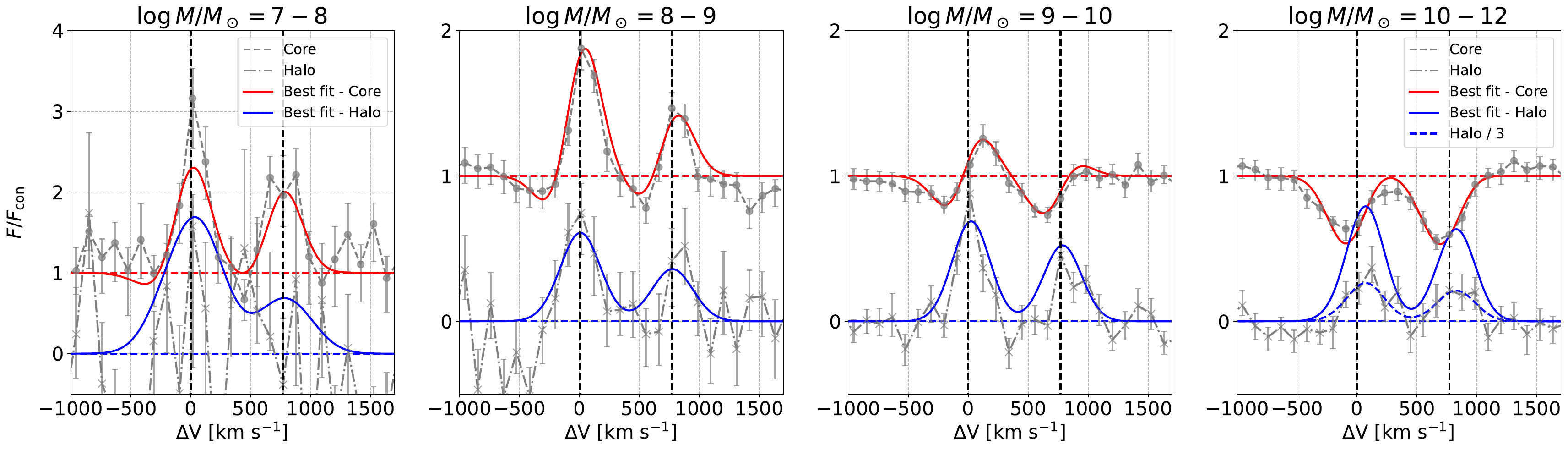}
    \includegraphics[width=\textwidth]{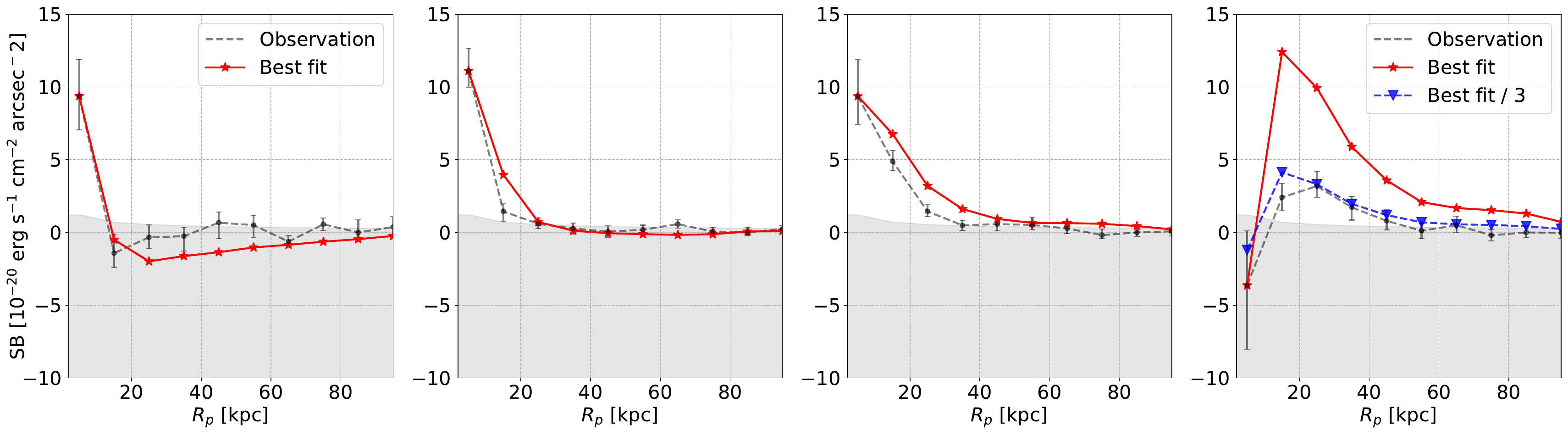}
    \caption{This figure presents the fitting of stacked spectra (top) and surface brightness (bottom) for four distinct stellar mass bins: $\log \mstar / \msun = 7-8$ (first), 8-9 (second), 9-10 (third), and 10-12 (fourth). 
    In the top panels, gray dashed and dot-dashed lines denote the stacked spectra in the core (\rp < 10 kpc) and halo (\rp = $10-30$ kpc) regions, respectively, corresponding to the spectra in Figure~\ref{fig:stack_spec}. The red solid lines illustrate the best-fit simulated spectra with the minimum $\chi^2$, based on the core spectrum analysis. The blue dashed lines represent the halo spectrum from the best model for the core spectrum. 
    In the bottom panels, the black dashed lines denoted the stacked surface brightness, corresponding to those in Figure~\ref{fig:stack_SB}.
    The red solid line illustrated the best-fit surface brightness from the best model for the core spectrum.
    For lower mass ranges ($\log \mstar / \msun < 9$), depicted in the top left panels, the simulated halo spectra align closely with the observed spectra. Conversely, in the right panel for the highest mass range ($\log \mstar / \msun = 10-12$), the simulated spectrum significantly overestimates the observed halo emission. The surface brightness of the best fit divided by three (blue dashed) is similar to the observed one.
    }
    \label{fig:fit_stack}
\end{figure*}

The resonant nature of the \mgii doublet photons allows for interactions with cold gas across a wide range of conditions, provided the \mgii column density (\Nmg) exceeds $10^{13} \unitNHI$, as shown in Figure~\ref{fig:spec_par}. These interactions result in \mgii absorption features from cold gas within star-forming regions or the ISM, P-Cygni profiles from \mgii scattering in outflowing gas, and spatially extended \mgii emission due to scattering of the gas in the outskirts of galaxies. Given that \mgii spectra from galaxies are often observed with the stellar continuum, the resultant spectral profiles encapsulate these phenomena alongside inherent \mgii absorption within the stellar continuum. Nonetheless, modeling based only on the core spectrum ($\rp < 10$ kpc) is insufficient to disentangle the various mechanisms contributing to the \mgii line formation. This is in the case of galaxies with \mstar/\msun $> 10^{10}$, where the fitted expansion velocities \vexp, predominantly cluster around $0 \kms$, indicating absorption features concentrated near the line center in this mass range, as shown in Figure~\ref{fig:2d_map}. This scenario complicates the differentiation of \mgii absorption attributable to either the gas within or around galaxies. This section extends the investigation to stacked spectra within the halo region ($\rp = 10-30$ kpc), aiming to clarify these distinctions.

\begin{table*}
\caption{Fitting parameters through the radiative transfer modeling of stacked spectra in the core region at $\rp < 10\, \rm kpc$ for four stellar mass bins.}
\centering
\begin{tabular}{llccccc}
\hline
       $\mstar/\msun$ &  & $\log \Nmg$ [\unitNHI]  &  \vexp [\kms] &  \sigran [\kms] &  \sigsrc [\kms] &  \EWint [\AA] \\ \hline 
\multirow{3}{*}{ $10^{7-8}$} 
& Average$^\ddag$ & 13.5 & 222 & 121 & 39 & 11.7 \\
& $\sigma^*$ & 0.49 & 197 & 57 & 19 & 3.7 \\
& Best$^\dag$ & 14 & 200 & 200 & 25 & 20 \\ \hline
\multirow{3}{*}{ $10^{8-9}$} 
& Average$^\ddag$    & 14.0  & 250   & 52   & 31    & 8.2 \\
& $\sigma^*$         & 0.12  & 51   & 32    & 11    & 1.6 \\
& Best$^\dag$        & 14    & 250   & 25   & 25    & 7 \\ \hline 
\multirow{3}{*}{ $10^{9-10}$} 
& Average$^\ddag$    & 15.2  & 120   & 28   & 46    & 4.3 \\
& $\sigma^*$         & 0.77  & 28   & 8.3    & 23    & 2.6 \\
& Best$^\dag$        & 14.5    & 150   & 25   & 50    & 4 \\ \hline 
\multirow{3}{*}{ $10^{10-12}$} 
& Average$^\ddag$    & 14.8  & 112   & 48   & 81    & 1.7 \\
& $\sigma^*$         & 0.39  & 35   & 17    & 54    & 2.8 \\
& Best$^\dag$        & 15    & 100   & 50   & 25    & 2 \\ \hline 
\end{tabular}
\\
\footnotesize{
$\ddag$: Weighted average in Eq.~\ref{eq:average}, 
$*$: Weighted standard deviation in Eq.~\ref{eq:error}.}
$\dag$: parameters of best-fit model with minimum $\chi^2$ in Eq.~\ref{eq:chisq}, 
\label{tab:par_stack}
\end{table*}

Figure~\ref{fig:fit_stack} shows the stacked spectra in both core and halo regions, as well as the SB profiles with the best-fit models with a minimum \chisq value for different mass bins. The best-fit parameters are shown in Table~\ref{tab:par_stack}.  
We found the best fit using the observed spectrum in the core region at $\rp < 10$ kpc. The best fits for the halo spectrum and the SB profiles are those of the model with parameters estimated by fitting the spectrum in the core region.
It can be seen from the top panels that the best-fit spectra are well-matched with the observed ones in the core region.
For low-mass galaxies (\mstar/\msun $< 10^{9}$), the halo spectra and SB profiles from the best-fit model closely match the observations. However, for higher masses (\mstar/\msun $> 10^{9}$), the model predicts stronger halo spectra and higher SB profiles compared to observations, with this discrepancy becoming especially pronounced for \mstar/\msun $> 10^{10-12}$.

The notable differences observed in the halo region for galaxies with \mstar/\msun $> 10^{10-12}$ indicate that intrinsic \mgii absorption in massive galaxies significantly affects \mgii emission in their outskirts. Additionally, anisotropic distributions of cold \mgii gas \citep[e.g.][]{guo23,pessa24} can affect the observed halo emission. In the right top panel of Figure~\ref{fig:fit_stack}, we illustrate the modified halo spectrum, which represents the original best fit reduced by a factor of three. Using this scaled model, we also recalculated the SB profile, which now closely matches the observed profiles. Therefore, we conclude that intrinsic absorption, anisotropic gas distribution, and dust in CGM significantly weaken the halo emission, reducing it by up to 50\%.

\subsection{Impact of Asymmetric Gas Distribution and Intrinsic Absorption on Surface Brightness Profiles}
\label{sec:bipolar}

\begin{figure*}
    \centering
    \includegraphics[width=\textwidth]{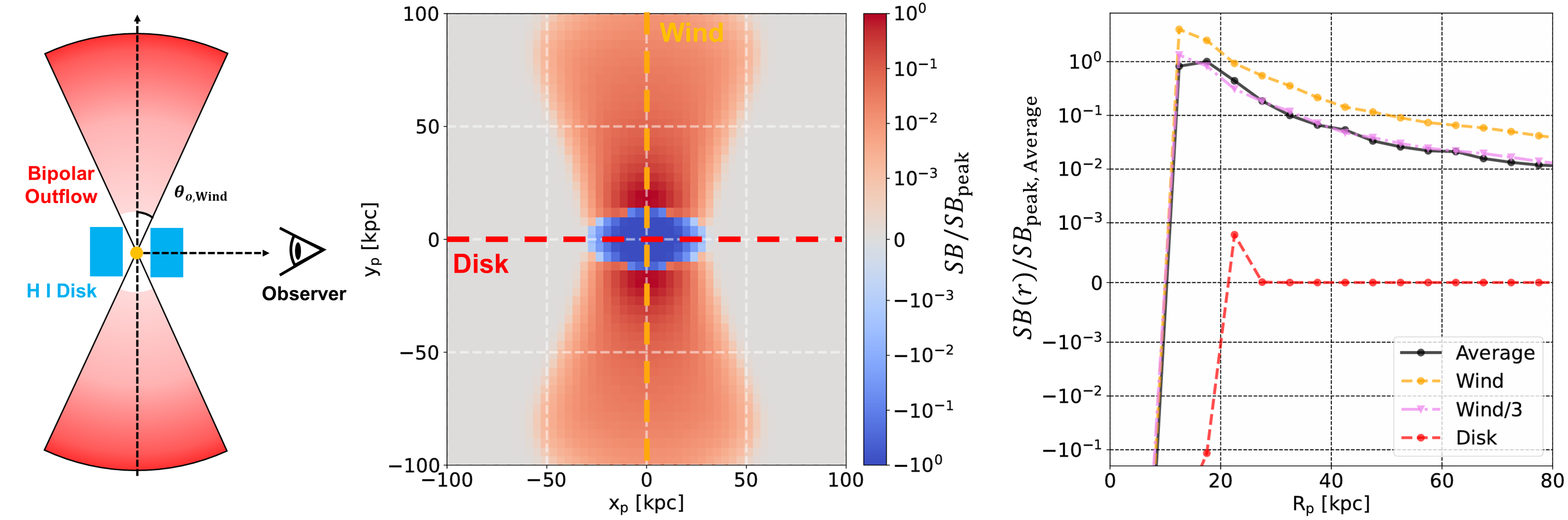}
    \caption{
 \mgii surface brightness (SB) map and profiles for the bipolar wind model. \textit{Left:} Schematic illustration of the bipolar wind model, including a central point source (orange), bipolar outflows (red), and a static \hi disk (blue). The line of sight is perpendicular to the bipolar wind, viewed edge-on. \textit{Center:} Simulated \mgii SB map with a spatial resolution of $\sim 6$ kpc. Red indicates \mgii emission, while blue shows absorption. \textit{Right:} \mgii SB profiles as a function of projected radius \rp. The orange and red dashed lines represent profiles along the `Wind' and `Disk' directions, respectively, as shown in the center panel. The black solid line is the azimuthally averaged profile (corresponding to an observed `stacked' profile), which closely matches the `Wind' direction profile divided by three (violet dashed line), indicating that asymmetric gas distribution reduces the overall SB. 
    }
    \label{fig:bipolar}
\end{figure*}

In this section, we explore the impact of asymmetric cold gas distribution and intrinsic absorption on the \mgii surface brightness (SB) profile. Figure~\ref{fig:bipolar} illustrates the schematic of the asymmetric wind model and its SB profiles. The wind model consists of a central point source, a bipolar outflow for asymmetric gas distribution, and a static inner \hi disk for intrinsic \mgii absorption. The outflow is characterized by an opening angle $\theta_{\rm o, Wind}$, sharing the same kinematics and gas distribution as the spherical geometry described in \S~\ref{sec:modeling}. These configurations allow us to assess how cold gas distribution influences the observed \mgii emission. We use the best-fit parameters for the stacked spectrum with \mstar/\msun $> 10^{10-12}$ from Table~\ref{tab:par_stack}. The \hi disk has an \mgii column density of $10^{15} \unitNHI$ along the equatorial direction, with fixed random motion and temperature at 100 \kms and $10^4\, \rm K$, respectively. The height and radius of the disk are 1 kpc and 5 kpc, respectively. We generate $10^8$ photons using a flat continuum with intrinsic \mgii emission (\EWint = 2 \AA) and collect photons, assuming the line of sight is perpendicular to the outflow direction.
 
The center panel of Figure~\ref{fig:bipolar} shows the simulated SB map. The \mgii feature appears as absorption at the center due to the optically thick HI disk, while emission is visible in the extended halo due to scattering within the wind. These combined absorption and emission features resemble the observed stacked spectra with \mstar/\msun $> 10^{10-12}$, as shown in the right panel of Figure~\ref{fig:fit_stack}, and also reported in other studies \citep{guo23}.

The right panel of Figure~\ref{fig:bipolar} compares the SB profiles along the wind (orange) and disk (red) directions. The wind profile is significantly more extended than that of the disk, indicating a directional influence of gas distribution. Naturally, for the stacked SB maps used here that do not take into account the orientation of the disk due to lack of high spatial resolution imaging, the averaged SB profile reflects the combined effect of different azimuthal angles.  The average profile (black) in the right panel of Figure~\ref{fig:bipolar} matches the `Wind' direction profile when reduced by a factor of three, highlighting that asymmetric gas distributions can significantly suppress the overall SB. As the opening angle decreases, this suppression factor increases, suggesting that the shape and directionality of the gas play a critical role in the observed emission.

Specifically, as demonstrated, the choice of $\theta_{\rm 0,wind}=30^\circ$ yields a suppression by a factor of $\sim 3$ -- just as observed (cf. \S~\ref{sec:modeling_halo}). Thus, the inclusion of effects due to asymmetric gas distributions represents an alternative, elegant explanation to dust for the lower SB observed. However, in order to fully compare to the observed stacked data set, we would need to loosen the assumption of the wind being orthogonal to the line-of-sight as done (implicitly) here. We will explore this in future work.

\section{ Conclusions }
\label{sec:conclusion}

In this study, we explored the properties of cold gas in and around galaxies at $z \sim 1$ by analyzing \mgii spectra through radiative transfer modeling and data stacking techniques. Using a comprehensive sample of 624 galaxies from the MUSE Analysis of Gas around Galaxies (MAGG) and the MUSE Ultra Deep Field (MUDF) programs, we focused on both core (0-10 kpc) and halo (10-30 kpc) regions to gain insights into the distribution and kinematics of cold gas. We successfully modeled 167 individual detections and four stacked datasets for different stellar mass bins, as well as \mgii emission extending beyond 10 kpc.

To do so, we assembled a suite of $>200,000$ radiative transfer models spanning the five-dimensional parameter space. Our main findings are as follows.
\begin{itemize}
    \item In spite of the simplicity of the model, we could reproduce complex observed \mgii spectra as well as surface brightness profiles. Specifically, only 3\% (5/167) of individual detections result in chi-square values $\chi^2$, higher than 10 (Appendix~\ref{sec:fitting}).

    \item Our analysis of stacked \mgii spectra for different stellar mass bins revealed significant variations in spectral profiles and surface brightness with stellar mass (\S~\ref{sec:stacking_observation}). In galaxies with lower stellar masses (\mstar/\msun $< 10^9$), \mgii emission was observed in both the core (\rp $< 10$ kpc) and halo regions (10 kpc $< \rp < 30$ kpc). Conversely, in higher mass galaxies (\mstar/\msun $> 10^{10}$), strong absorption features dominated in the core, while the \mgii surface brightness was more extended. This trend indicates that more massive galaxies have abundant cold CGM, leading to mass-dependent variations in cold gas distribution.
    
    \item Radiative transfer modeling allowed us to explore the relationships among key fitting parameters, such as the Mg~II column density \Nmg and expansion velocity \vexp (Figure~\ref{fig:2d_map} in \S~\ref{sec:fitting_result}). We found a negative correlation between \Nmg and \vexp, indicating that regions with higher column densities tend to contain slower-moving gas. The expansion velocity was generally found to cluster around zero and 200 \kms, suggesting the coexistence of static or slowly moving gas and outflows. Moreover, these parameters varied depending on the spectral type—`Absorption,' `P-Cygni,' and `Emission'—revealing different physical conditions of the cold gas. Notably, higher stellar mass galaxies exhibited higher \Nmg and lower \vexp values, indicating an abundance of slowly moving cold gas in massive galaxies.

    \item The stacking analysis allowed us to access and model not only the core but also the halo spectra, thus allowing us to directly constrain the physical conditions of the CGM through emission (\S~\ref{sec:modeling_halo}). 
    By capturing simulated photon packages escaping in different projected radii bins, we could fit the stacked `core' and `halo' spectrum simultaneously. While the overall fit of both spectra is very good, we overproduce the `halo' spectrum of the highest mass bin ($\mstar/\msun > 10^{10}$) by a factor $\sim 3$. 
    
    \item 
    Asymmetric gas distribution significantly reduces the stacked \mgii surface brightness, with directional gas flows leading to notable variations in observed profiles (\S~\ref{sec:bipolar}). This finding highlights the importance of accounting for anisotropic gas when interpreting extended \mgii emission, particularly in the CGM of massive galaxies.
\end{itemize}

Interpreting observations of resonant lines such as \mgii are notoriously complex due to the non-linear radiative transfer effects shaping the emergent observables. However, this process also holds the potential to probe the non-intrinsically emitting gas found, for instance, in the outskirts of galaxies. 
Advances in theoretical insight and computing power allow us to fit complex, observed spectra using full radiative transfer models and, thus, to obtain physical insights from resonant lines. 
In this work, we showed that it is possible to successfully reproduce a large number of observed spectra using this approach and obtain physical interpretable parameters. We furthermore correlate these parameters with galactic properties such as the stellar mass, and reveal several (anti-)correlations between them.


As several additional parameters, such as the asymmetric gas distribution, the dust content or related to the multiphase nature of the scattering medium, are not captured in our fitting pipeline, we see this effort as a first step in an ongoing process and plan to revisit this problem to better address the growing amount of resonant emission line data available.

\section*{Acknowledgements}
MG thanks the Max Planck Society for support through the Max Planck Research Group. Computations were performed on the HPC system Freya and Orion at the Max Planck Computing and Data Facility. This work is based on observations collected at the European Organisation for Astronomical Research in the Southern Hemisphere under ESO programme IDs 197.A-0384 and 1100.A-0528.


\section*{Data Availability}
Data related to this work will be shared on reasonable request to the corresponding author.
The MUSE data used in this work are available from the European Southern Observatory archive (\url{https://archive.eso.org}).



\bibliographystyle{mnras}
\bibliography{mybib} 



\appendix

\section{Deatils of the \mgii spectral fitting pipeline}
\label{sec:fitting}

\begin{figure*}
    \centering
    \includegraphics[width=\textwidth]{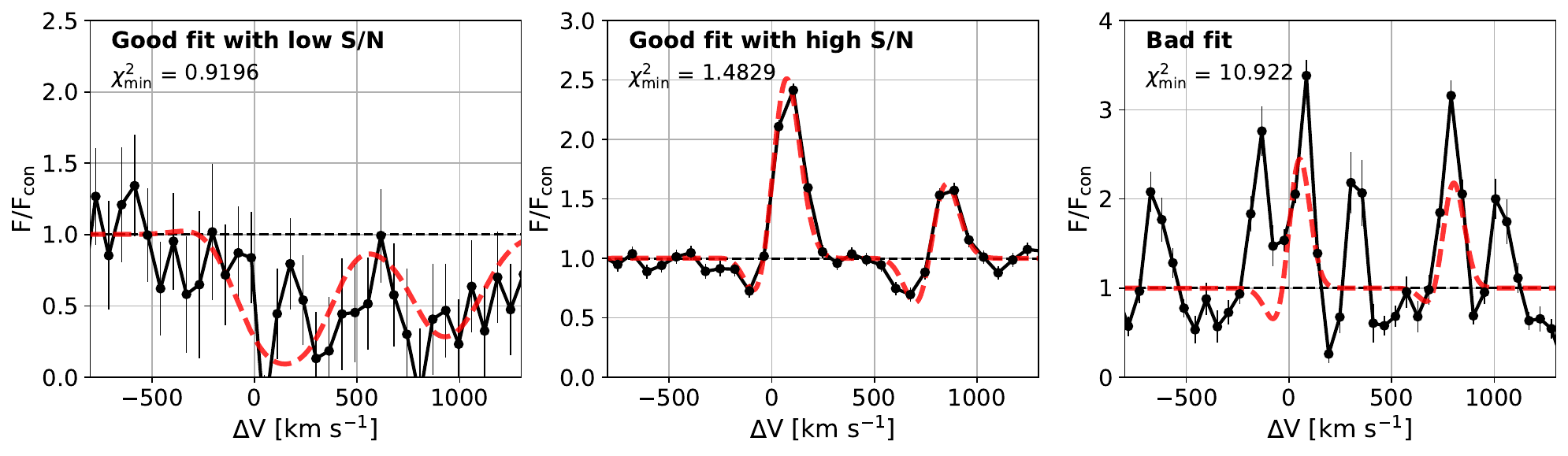}
    \caption{
    Best fitting results of three individual spectra.
    The left and center panels show 'good' fit examples of spectra with low and high S/N, respectively.
    The right panel shows a 'bad' fit example for $\chi_{\rm min} > 10$.
    The black and red lines represent the observed and simulated spectra at the minimum $\chi^2$.
    }
    \label{fig:fit_indi}
\end{figure*}

\begin{figure}
    \centering
    \includegraphics[width=0.5\textwidth]{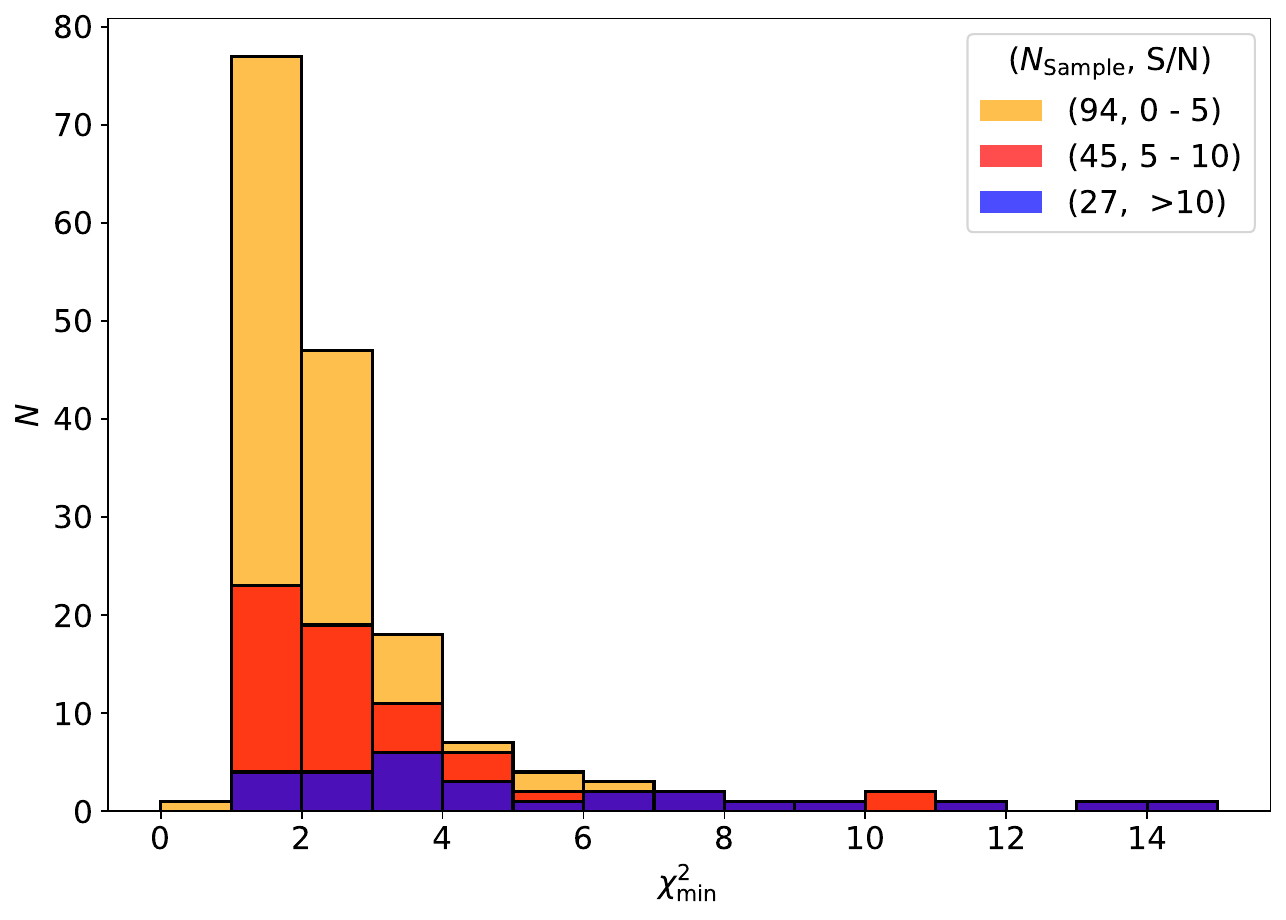}
    \caption{$\chi^2_{\rm min}$ histogram of individual detections. Colors represent the range of \mgii signal-to-noise, 0-5 (orange), 5-10 (red), > 10 (blue).}
    \label{fig:histogram_chi}
\end{figure}

\begin{figure*}
    \centering
    \includegraphics[width=\textwidth]{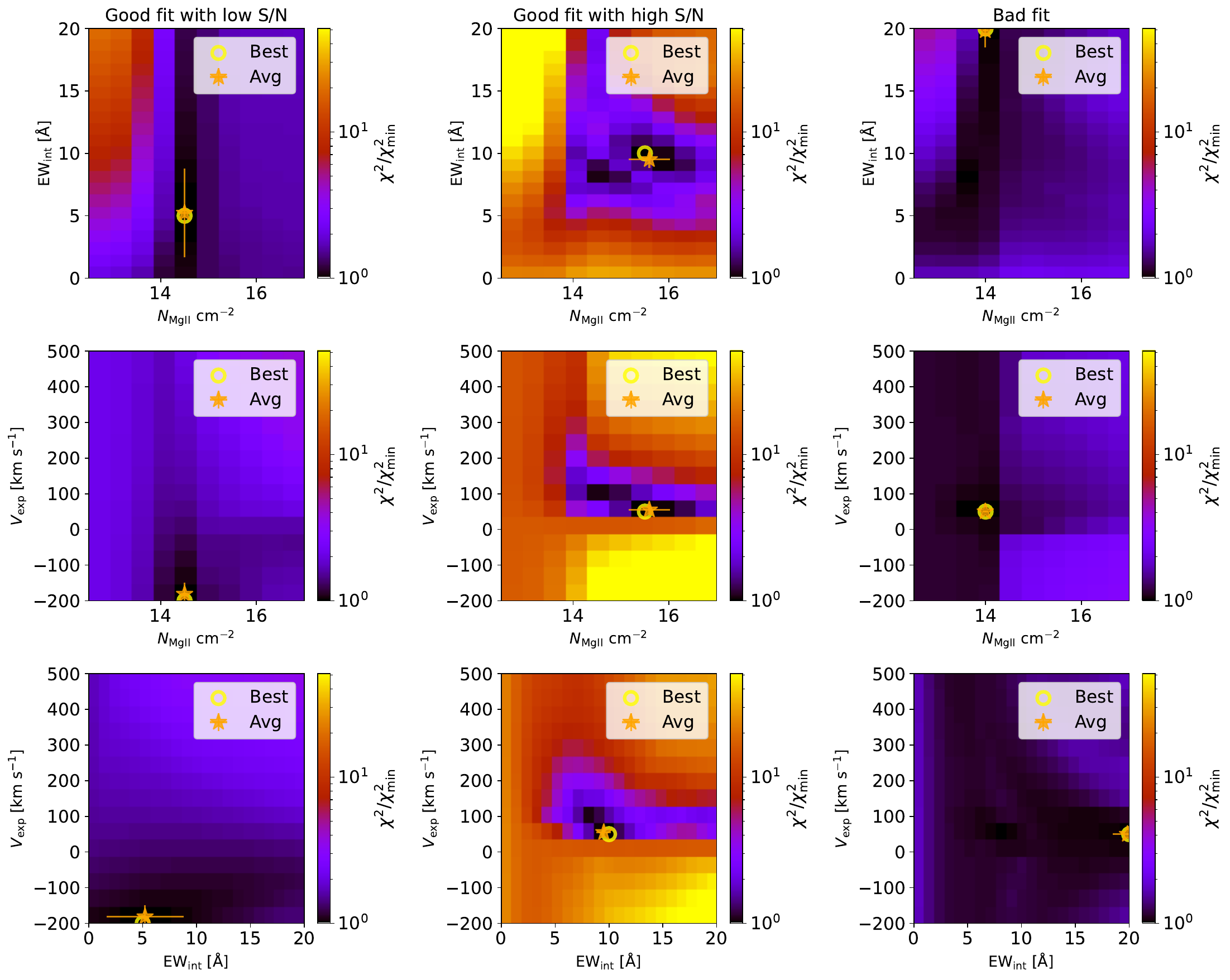}
    \caption{ $\chi^2$ maps in \EWint - \vexp (top) and \EWint - \NHI (bottom) planes of three spectra in Figure~\ref{fig:fit_indi}. Each $\chi^2$ value is the minimum as two parameters in the $x-y$ axis are fixed.
    The yellow star marks represent the parameters of the best fit at the minimum $\chi^2$. The red open circles represent the values of the weighted average in Eq.~\ref{eq:average} with the error bar corresponding to the weighted standard deviations in Eq.~\ref{eq:error}.
    }
    \label{fig:chi_map}
\end{figure*}

\begin{figure*}
    \centering
    \includegraphics[width=\textwidth]{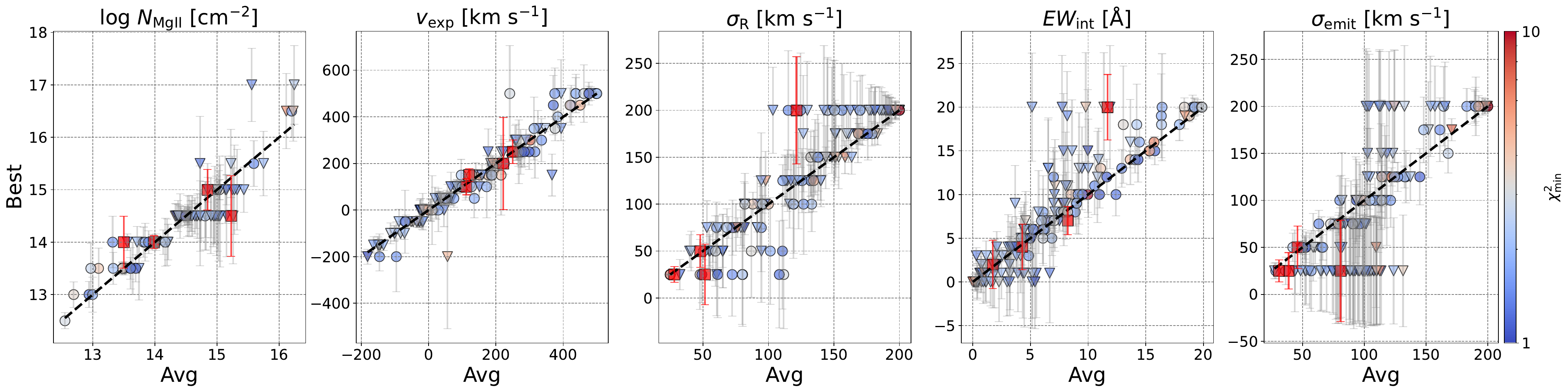}
    \caption{Comparisions of the parameters, \NHI, \vexp, \sigr, \EWint, \sigsrc between the weighted average and the best fit, which are $x$ and $y$ axes, respectively. The blue circles and inverse triangles represent individual spectra classified as 'emission' \& 'P-Cygni' and 'absorption,' respectively, as shown in Figure~\ref{fig:obs_prop}.
    The black dashed line is the $x=y$ graph. The error bar means the weighted standard division. Most individual and stacking spectra show similar weighted averages and best fit parameters.
    }
    \label{fig:best_avg}
\end{figure*}

This section describes how we analyze and fit the observed \mgii spectra using simulated spectra to discern the scattering signature within the cold medium. 
We use a chi-squared test to fit the observed spectra with the simulated counterparts.
For a comparative analysis of the two spectra, we collect the simulated spectrum in a radius of 0 to 10 kpc as \mgii spectra of individual objects are observed in this radius range. We also adopt the typical MUSE spatial resolution of $\sim 6$ kpc at $z \approx 1$ through 2D Gaussian convolutions.
Furthermore, we consider the observed spectral resolution.
Because the spectral resolution of MUSE varies as a function of wavelength \citep{Weilbacher2020}, the spectral resolution variation must be considered to compare the observed and simulated spectra.
The MUSE spectral resolution\footnote{see the MUSE spectral resolution as a function of wavelength at the webpage: https://www.eso.org/sci/facilities/paranal/instruments/muse/inst.html} can be reproduced using the below relation:
\begin{equation}\label{eq:MUSE_resolution}
    R(\lambda) = p_0 + p_1 \lambda p_2 \lambda^2 + p_3\lambda^3,
\end{equation}
where $p_0 = 499.446$, $p_1 = -0.201608$, $p_2 = 1.3029 \times 10^{-4}$, and $p_3 =7.87479 \times 10^{-9}$, and $\lambda$ is in units of \AA.
In the fitting process, the spectral resolution $R_{\rm obs}$ is $\sim R(\lambda = (\zobs+1) \times 2800\textup{~\AA})$ as the wavelength of \mgii doublet is in the rest frame $\sim 2800$ \AA.
We convolve the simulated spectra with the Gaussian function at the width $\sim 2.355 R_{\rm obs}$ as $R(\lambda)$ represents the full-width half maximum of the spectral resolution.
In the \zobs range of our samples (from 0.6 to 2.2), the spectral resolution is $1500-3500$, corresponding to the Gaussian width of 35-85 \kms; $R(\lambda)$ increases with increasing \zobs.
Note that $R = 1000$, which is half of the average spectral resolution, is adopted to compare the simulated spectra with the stacked spectra, as the stacking process can lead to broadening of the line due to uncertainty in the redshifts, thus effectively degrading the resolution.
After that, we normalize the observed spectrum by dividing it by the continuum level, which is obtained by averaging the spectrum near $-1500$ \kms from \mgii $\lambda$2796. The normalized flux of the continuum is fixed at 1 like in the simulated spectra in Figure~\ref{fig:spec_par}. Also, the observed wavelength is converted to the wavelength in the rest frame by dividing by (1+$z_{\rm obs}$).

Given that the normalized observed spectrum and the mock simulated spectrum are \fsim and \fobs,
we estimate the chi-square value $\chi^2$ given by
\begin{equation}\label{eq:chisq}
\chi^2 = {{ \sum^{N_{\rm obs}}_i \frac{(\fobs (\lambda_i) - \fsim (\lambda_i))^2}{ \sigma_{\rm obs}^2 (\lambda_i) }} / {N_{\rm obs}}},
\end{equation}
where $\lambda_i$ is the wavelength in the rest frame and $\sigma_{\rm obs}$ is the error of the observed spectrum.
$N_{\rm obs}$ is the number of data points of the observed spectra in the wavelength range from 2788 \AA\ to 2808 \AA\, which is 
the selected velocity range, from $-800$ to $+1300$ \kms relative to \mgii $\lambda$2796, capturing relevant features for our analysis.
We estimate $\chi^2$ values of all simulated \mgii spectra.
Consequently, the simulated spectrum is defined as the best fit when the chi-square value $\chi^2$ is minimized, providing an optimal representation of the observed spectra through our fitting process.

The minimum chi-square value $\chi^2_{\rm min}$ depends on the signal-to-noise of \mgii $S/N$.
Figure~\ref{fig:fit_indi} shows fitting examples for various $\chi^2_{\rm min}$. 
In the left and center panels of Figure~\ref{fig:fit_indi},
we show examples of two cases with low S/N ($\sim 2$) and high S/N ($\sim 17$), respectively.
The fitting quality in the center panel is better than that in the left panel, although its $\chi^2_{\rm min}\sim 0.9$ is higher than in the left panel, $\sim 1.5$. 

The best fit is not always similar to an observed spectrum.
We show an example in the right panel of Figure~\ref{fig:fit_indi}, where the simulated spectrum does not match the observed spectrum because of multiple emission peaks ($\chi^2_{\rm min} > 10$).
Figure~\ref{fig:histogram_chi} shows the histogram of the minimum chi-square values $\chi^2_{\rm min}$ for various signal-to-noise $S/N$.
The $\chi^2_{\rm min}$ of only five fit is higher than 10.

Figure~\ref{fig:chi_map} displays the $\chi^2$ maps of three spectra from Figure~\ref{fig:fit_indi}. These maps are presented in the $\Nmg-\EWint$, $\Nmg-\vexp$, and $\EWint-\vexp$ planes.
In the top central panel of Figure~\ref{fig:chi_map}, values of the lower \Nmg and \EWint cases are comparable to the minimum.
This local minimum region causes a degeneracy in the fitting results. 
For example, if both \Nmg and \EWint increase in the best-fit model, a higher \Nmg suppresses \mgii emission feature and a higher \EWint enhances the feature conversely (see Figure~\ref{fig:spec_par}). 
In this case, the simulated spectrum with higher \Nmg and \EWint becomes similar to the best fit due to the opposite influence of the two parameters.

To avoid these degeneracies,
we compute the weighted average of each simulated parameter.
The weighted average of \Nmg is given by
\begin{equation}\label{eq:average}
\langle\Nmg\rangle = \sum^{N_{\rm Sim}}_j N_{\rm Mg II, j}  \exp \left( -N_{\rm obs} \frac{\chi_j^2 }{ 2} \right),
\end{equation}
where $N_{\rm Sim}$ is the number of simulated spectra (201,600). Each simulated spectrum has a corresponding $\chi^2$ value denoted as $\chi_j^2$.
Furthermore, the weighted standard deviation of \Nmg
\begin{equation}\label{eq:error}
\sigma_\Nmg = \langle\Nmg^2\rangle - \langle\Nmg\rangle^2 ,
\end{equation}
where $\langle\Nmg^2\rangle$ is the weighted average of $\Nmg^2$.
In Figure~\ref{fig:chi_map}, the orange star marks for the weighted average exist near points of the yellow open circle for the best-fit parameter.

To check the correlation between parameters of best fit and weighted average, we compare two values in Figure~\ref{fig:best_avg}.
Apart from a few objects, the two values are in a strong linear correlation $y=x$ in most individual spectra.
In the fourth and fifth panels, some objects have large error bars (the weighted standard deviation) because these objects do not have enough emission to measure \EWint and \sigsrc; the large error bar is along the \EWint axis in the left panels of Figure~\ref{fig:chi_map}.

In summary, 
we fit the observed \mgii spectra using simulated spectra generated by \texttt{RT-scat}.
We estimate the $\chi^2$ values using 201,600 
simulated spectra and select as the best-fit model the one where $\chi^2$ is minimum.
After that, we calculate the weighted average and standard deviation for each parameter to avoid fitting degeneracies. 
As a result, we set the fitting parameters as the weighted average and explore the trends of these parameters. 

\bsp	
\label{lastpage}
\end{document}